\documentclass[12pt,a4paper]{article}

\usepackage{latexsym}
\usepackage{amsfonts}
\usepackage{amssymb}
\usepackage{amsmath}
\usepackage{graphicx}
\usepackage[dvips]{psfrag}
\usepackage{vmargin}
\usepackage{cite}

\setpapersize{A4}
\setmargins{25mm}{20mm}{160mm}{220mm}{12pt}{20pt}{12pt}{36pt}

\newcommand\cmp[3]  {{\it Comm.\ Math.\ Phys.\ }{\bf #1} (#2) #3}

\newcommand\jhep[3] {{\it J. High Energy Phys.\ }{\bf #1} (#2) #3}

\newcommand\npb[3]  {{\it Nucl.\ Phys.\ }{\bf B #1} (#2) #3}

\newcommand\pla[3]  {{\it Phys.\ Lett.\ }{\bf A #1} (#2) #3}
\newcommand\plb[3]  {{\it Phys.\ Lett.\ }{\bf B #1} (#2) #3}

\newcommand\prd[3]  {{\it Phys.\ Rev.\ }{\bf D #1} (#2) #3}

\newcommand\ptp[3]  {{\it Prog.\ Theor.\ Phys.\ }{\bf #1} (#2) #3}
\newcommand\rmp[3]  {{\it Rev.\ Mod.\ Phys.\ }{\bf #1} (#2) #3}

\newcommand{\hepth}[1]{{\tt hep-th/#1}}

\newcommand{\scsc}{\scriptscriptstyle}
\renewcommand{\sc}{\scriptstyle}
\newcommand{\R}{\mathbb{R}}

\newcommand{\bz}{\bar{z}}
\newcommand{\htn}{\hat{n}}
\newcommand{\alb}{\bar{\alpha}}
\newcommand{\cd}{c^{\dag}}
\newcommand{\bra}[2]{\left<#1,#2\right|}
\newcommand{\cket}[2]{\left|#1,#2\right>}
\newcommand{\Deltah}{\hat{\Delta}}
\newcommand{\nh}{\hat{n}}
\newcommand{\Nh}{\hat{N}}

\begin{document}

\begin{titlepage}

\begin{flushright}
 HUPD-0117\\
 \hepth{0201196}
\end{flushright}

 \vspace*{10mm}

 \begin{center}

  {\LARGE\sffamily\bf Instanton Number
  Calculus\vspace{2mm}\\on\vspace{3mm}\\
  Noncommutative $\R^4$}
  \vspace{25mm}\\

  {\sffamily\bf Tomomi Ishikawa,\footnote{{\it E-mail} :
  \ttfamily tomomi@theo.phys.sci.hiroshima-u.ac.jp}
  Shin-Ichiro Kuroki\footnote{{\it E-mail} :
  \ttfamily kuroki@theo.phys.sci.hiroshima-u.ac.jp} and
  Akifumi Sako\footnote{{\it E-mail} :
  \ttfamily sako@math.sci.hiroshima-u.ac.jp}\vspace{15mm}}\\
  {\it Graduate School of Science, Hiroshima University,\\
  1-3-1 Kagamiyama, Higashi-Hiroshima 739-8526, Japan}\vspace{20mm}\\

 \end{center}


 \begin{center}
  {\bf ABSTRACT}
 \end{center}

 In noncommutative spaces, it is unknown whether the
 Pontrjagin class gives integer, as well as, the relation between the
 instanton number and Pontrjagin class is not clear.
 Here we define ``Instanton number'' by the size of $B_{\alpha}$
 in the ADHM construction.
 We show the analytical derivation of the noncommuatative $U(1)$
 instanton number as an integral of Pontrjagin class (instanton charge)
 with the Fock space representation. Our approach is for the arbitrary
 converge noncommutative $U(1)$ instanton solution, and is based on
 the anti-self-dual (ASD) equation itself.
 We give the Stokes' theorem for the number operator representation.
 The Stokes' theorem on
 the noncommutative space shows that instanton charge is given by some
 boundary sum. Using the ASD conditions, we conclude that
 the instanton charge is equivalent to the instanton number.


\end{titlepage}

\section{Introduction}

Recently, there has been much interest in noncommutative field theory
motivated by the string theory
 \cite{Connes2,Douglas1,Seiberg2,Douglas2,Okawa1}.
For example, the noncommutative gauge theory arises on D-branes
in the presence of background constant Neveu-Schwarz $B$ field.
The discoveries show us the analysis of noncommutative gauge theory is
very important for nonperturbative analysis of the string theory.
At the same time noncommutative instantons are one of the great
interests for many physicists, since
the instanton plays an important role of the
nonperturbative analysis of the Yang-Mills theory.

In commutative space, there is a well-known method to construct
instanton solution, which is given by Atiyah, Drinfeld, Hitchin and
Manin (ADHM)~\cite{Atiyah1,Corrigan1,Osborn1}.
There is the one-to-one
correspondence between the instanton solutions and the ADHM data.
On the other hand, in noncommutative spaces case, a pioneering work for
instantons was done by Nekrasov and Schwarz \cite{Nekrasov1}
(and see also
\cite{Furuuchi1,Nekrasov2,Lee1,Chu1,Kim1,Kim2,Correa1,Lechtenfeld1,Kraus1}).
They showed that noncommutative instanton solutions are obtained
by deformed ADHM equations, where the deformation of
the equations is caused by the noncommutativity of spaces.
One of the remarkable feature of instantons on noncommutative spaces is
their including some kind of resolution of singularities.
For example, there is no $U(1)$ instanton in commutative spaces
because of small instanton singularities.
On the contrary in noncommutative spaces the point of spacetime is
blurred, and the singularities are smeared, therefore instantons exist.

We constructed the elongated type $U(1)$ instanton solution for
arbitrary instanton number \cite{Ishikawa1}.
(This solution is constructed by the same ADHM data as Braden and
Nekrasov used in \cite{Braden1}, that is the elongated instanton in
``commutative'' spaces but with nontrivial metric.)
Here we define ``instanton number'' by the
${\rm dim}~V$ of $B_{\alpha}\in Hom(V,~V)$ that appear in
ADHM data and $V$ is a complex vector space i.e.$V={\bf C}^k$.
In other words, we define ``instanton number'' by a rank
of projection~\cite{Furuuchi1}. In commutative case, instanton number is
given by the Pontrjagin class and it is equivalent to ${\rm dim}~V$.
However in noncommutative spaces, many problems are left for the defining
of the integer-valued pontryagin class, as well as for the proof of the
identity between the instanton number and the pontryagin class.
There are some explanations about the relation between the
Pontrjagin class and the instanton number (\cite{Furuuchi1,Chu1} and so on).
They say the integral of the Pontrjagin class
is equivalent to the $ dim (1-P) $ and instanton number. 
(Here $P$ is a projector that
appears in the ADHM construction.)
As we will see, this result is exactly same as the result of this article.
But their explanations are not strict proof because these discussions
contain some gaps and calculations performed including infinity.
For example, trace operation has no cyclic symmetry in noncommutative
theories
(total divergence terms appear when order of operators is changed 
by the cyclic rotating).
When we estimate topological charge like the integral of the Pontrjagin
class, the surface terms are essential.
So we have to estimate carefully all surface terms including 
the terms caused by using cyclic rotating in trace operation.
This problem is not solved by the pure gauge condition.
But there has been no proof that estimate all surface terms.
Another type of the problem is infinity.
For example, strictly speaking the result $dim (1-P) $ is not
defined  because it takes arbitrary value.
If we try to estimate them carefully without ambiguity from infinity,
 there is only way that we introduce some cut-off
(or boundary) and calculate all the surface terms.
But there has been no proof containing such approach. 
So we give it in this article.
All our calculations are done in finite and all the
surface terms are estimated.

In \cite{Ishikawa1} we showed that
 the elongated type $U(1)$ instanton solution gives a
integer as a Pontrjagin number that is equal to instanton number, by
numerical way. We call instanton charge as the integral of the
Pontrjagin class in this article to distinguish it from instanton number.

In this article, we perform the calculation of the integral of the
Pontrjagin class analytically, for any instanton solution with some
converge condition. The elongated type $U(1)$ instantons belong to the
class of instantons that shrink to the origin and its projection operator
is diagonalized in the number operator representation consequently.
In this case, the instanton number calculus is relatively easy.
However, the case that the projection operator is not diagonalized
needs some techniques. We note that the cut-off, which we will
introduce and finally take infinite limit, plays an important role of
the Pontrjagin class calculus.\\

The organization of this paper is as follows.
In section 2, we review the noncommutative instanton and prepare
tools of instanton charge calculus.
In section 3, instanton charge is calculated in the elongated type
instanton case. This calculation is a simple example of the general
instanton charge  calculus.
In section 4, the general type of instanton charge is obtained under
some converge condition. Stokes' theorem of the Fock space formalism is
discussed, too.
In section 5, we summarize this article.

\section{Noncommutative $U(1)$ Instantons}
We summarize the noncommutative field theory
used in this article and review the
noncommutative instantons in this section.
\subsection{Noncommutative $\bf{R}^4$ and the Fock space representation}
Let us consider Euclidean noncommutative $\bf{R}^4$, whose coordinate
functions $x^{\mu}\;(\mu=1,2,3,4)$ on the deformed noncommutative
manifold satisfy the following commutation relations
\begin{equation}
 [x^{\mu},x^{\nu}]=i\theta^{\mu\nu}, \label{EQ:xx-commu}
\end{equation}
where $\theta^{\mu\nu}$ is an antisymmetric real constant matrix,
whose elements are called noncommutative parameters.
We can always bring $\theta^{\mu\nu}$ to the skew-diagonal form
\begin{equation}
 \theta^{\mu\nu}=
  \begin{pmatrix}
   0            & \theta^{12} & 0            & 0 \\
   -\theta^{12} & 0           & 0            & 0 \\
   0            & 0           & 0            & \theta^{34} \\
   0            & 0           & -\theta^{34} & 0
  \end{pmatrix}
\end{equation}
by space rotation.
For simplicity, we restrict the noncommutativity of the space to the
self-dual case of $\theta^{12}=\theta^{34}=-\zeta\;(\zeta>0)$.
Here we introduce complex coordinates
\begin{equation}
 z_1=\frac{1}{\sqrt{2}}(x^1+ix^2),\;\; z_2=\frac{1}{\sqrt{2}}(x^3+ix^4),
  \label{EQ:z-coordinates}
\end{equation}
then the commutation relations (\ref{EQ:xx-commu}) become
\begin{equation}
 [z_1,\bz_1]=[z_2,\bz_2]=-\zeta,\;\;\mbox{others are zero}.
  \label{EQ:z-commu}
\end{equation}
For using the usual operator representation, we define creation and
annihilation operators by
\begin{equation}
 \cd_{\alpha}=\frac{z_{\alpha}}{\sqrt{\zeta}},\;\;
  c_{\alpha}=\frac{\bz_{\alpha}}{\sqrt{\zeta}},\;\;
  [c_{\alpha},\cd_{\alpha}]=1\;\;\;\;(\alpha=1,2).\label{EQ:c-a-operator}
\end{equation}
The Fock space $\cal H$ on which the creation and annihilation
operators (\ref{EQ:c-a-operator}) act is spanned by the Fock state
\begin{equation}
 \left|n_1,n_2\right>
  =\frac{(\cd_1)^{n_1}(\cd_2)^{n_2}}{\sqrt{n_1!n_2!}}\left|0,0\right>,
\end{equation}
with
\begin{eqnarray}
 c_1\cket{n_1}{n_2}=\sqrt{n_1}\cket{n_1-1}{n_2},\;\;&&
  \cd_1\cket{n_1}{n_2}=\sqrt{n_1+1}\cket{n_1+1}{n_2},\nonumber\\
 c_2\cket{n_1}{n_2}=\sqrt{n_2}\cket{n_1}{n_2-1},\;\;&&
  \cd_2\cket{n_1}{n_2}=\sqrt{n_2+1}\cket{n_1}{n_2+1},
\end{eqnarray}
where $n_1$ and $n_2$ are the occupation number.
The number operators are also defined by
\begin{equation}
 \nh_{\alpha}=\cd_{\alpha}c_{\alpha},\;\;\Nh=\nh_1+\nh_2,
\end{equation}
which act on the Fock states as
\begin{equation}
 \nh_{\alpha}\cket{n_1}{n_2}=n_{\alpha}\cket{n_1}{n_2},\;\;
  \Nh\cket{n_1}{n_2}=(n_1+n_2)\cket{n_1}{n_2}.
\end{equation}
In the operator representation, derivatives of a function
$f(z_1,\bz_1,z_2,\bz_2)$ are defined by
\begin{eqnarray}
 \partial_{\alpha}f(z)=[\hat{\partial}_{\alpha},f(z)],\hspace{5mm}
  \partial_{\bar{\alpha}}f(z)=[\hat{\partial}_{\bar{\alpha}},f(z)],
\end{eqnarray}
where $\hat{\partial}_{\alpha}=\bar{z}_{\alpha}/\zeta$,
$\hat{\partial}_{\bar{\alpha}}=-z_{\alpha}/\zeta$ which satisfy
\begin{equation}
 [\hat{\partial}_{\alpha},\hat{\partial}_{\bar{\alpha}}]=-\frac{1}{\zeta}.
\end{equation}
The integral on noncommutative $\bf{R}^4$ is defined by the standard
trace in the operator representation,
\begin{equation}
 \int d^4x=\int d^4z=(2\pi\zeta)^2\mbox{Tr}_{\cal H}.
\end{equation}
Note that $\mbox{Tr}_{\cal H}$ represents the trace over the Fock
space whereas the trace over the gauge group is denoted by
$\mbox{tr}_{U(N)}$.

\subsection{Noncommutative gauge theory and instantons}
Let us  consider the $U(N)$ Yang-Mills theory on noncommutative $\bf R^4$.

In the noncommutative space, the Yang-Mills connection is defined by
\begin{equation}
 \hat{\nabla}_{\mu}\Psi=-\Psi\hat{\partial}_{\mu}+\hat{D}_{\mu}\Psi,
\end{equation}
where $\Psi$ is a matter field and $D_{\mu}$ are anti-hermitian gauge
fields \cite{Nekrasov2}\cite{Nekrasov3}\cite{Gross2}.
Then the Yang-Mills curvature of the connection $\nabla_{\mu}$ is
\begin{equation}
 F_{\mu\nu}=[\hat\nabla_{\mu},\hat\nabla_{\nu}]
  =-i\theta_{\mu\nu}+[{\hat D}_{\mu},{\hat D}_{\nu}].\label{EQ:F-x}
\end{equation}
In our notation of the complex coordinates (\ref{EQ:z-coordinates}) and
(\ref{EQ:z-commu}), the curvature (\ref{EQ:F-x}) is
\begin{equation}
 F_{\alpha\alb}=\frac{1}{\zeta}+[{\hat D}_{\alpha},{\hat D}_{\alb}],
  \hspace{5mm}
  F_{\alpha\bar{\beta}}=[{\hat D}_{\alpha},{\hat D}_{\bar{\beta}}]
  \hspace{10mm}(\alpha\not=\beta).
\end{equation}
The Yang-Mills action is given by
\begin{equation}
 S=-\frac{1}{g^2}{\rm Tr}_{\cal H} \mbox{tr}_{U(N)}F\wedge
*F,\label{EQ:Y-M-action}
\end{equation}
where we denote ${\rm tr}_{U(N)}$ as a trace for the gauge group U(N),
$g$ is the Yang-Mills coupling and $*$ is Hodge-star.

Then the equation of motion is
\begin{equation}
 [\nabla_{\mu},F_{\mu\nu}]=0.\label{EQ:Y-M-eom}
\end{equation}
(Anti-)instanton solutions are special solutions of (\ref{EQ:Y-M-eom})
which satisfy the (anti-)self-duality ((A)SD) condition
\begin{equation}
 F=\pm*F.\label{EQ:solution-original}
\end{equation}
These conditions are rewritten in the complex coordinates as
\begin{eqnarray}
 F_{1\bar{1}}=&+&F_{2\bar{2}},\;\;F_{1\bar{2}}=F_{\bar{1}2}=0\;\;\;\;
  \mbox{(self-dual)},\label{EQ:SD}\\
 F_{1\bar{1}}=&-&F_{2\bar{2}},\;\;F_{12}=F_{\bar{1}\bar{2}}=0\;\;\;\;
  \mbox{(anti-self-dual)}.\label{EQ:ASD}
\end{eqnarray}
In the commutative spaces, solutions of Eq.(\ref{EQ:solution-original})
are classified by the topological charge
(integral of the Pontrjagin class)
\begin{equation}
 Q=-\frac{1}{8\pi^2}\int\mbox{tr}_{U(N)}F\wedge F,\label{EQ:Q}
\end{equation}
which is always integer and called instanton number $k$.
However, in the noncommutative spaces above statement is unclear.
We discuss this issue in this article by using the operator
representation of (\ref{EQ:Q})
\begin{equation}
 Q=\begin{cases}
    \zeta^2\mbox{Tr}_{\cal H}\,\mbox{tr}_{U(N)}
    (F_{1\bar{1}}F_{2\bar{2}}-F_{12}F_{\bar{1}\bar{2}})\;\;\;\;&
    \mbox{(self-dual)}\\
    \zeta^2\mbox{Tr}_{\cal H}\,\mbox{tr}_{U(N)}
    (F_{1\bar{1}}F_{2\bar{2}}-F_{1\bar{2}}F_{2\bar{1}})\;\;\;\;&
    \mbox{(anti-self-dual)}.
    \end{cases}\label{EQ:Q-op-N}
\end{equation}
\subsection{Nekrasov-Schwarz noncommutative $U(1)$ instantons}
In the ordinary commutative spaces, there is a well-known way to find
ASD configurations of the gauge fields.
It is ADHM construction which is proposed by Atiyah, Drinfeld, Hitchin
and Manin \cite{Atiyah1}.
Nekrasov and Schwarz first extended this method to noncommutative
cases \cite{Nekrasov1}. Especially $U(1)$ cases are discussed in
\cite{Nekrasov2}\cite{Nekrasov3}\cite{Furuuchi1}\cite{Furuuchi2}
in detail. Here we review briefly on the ADHM construction of $U(1)$
instantons\cite{Nekrasov2}\cite{Nekrasov3}.

The first step of ADHM construction on noncommutative $R^4$ is looking
for matrices
$B_1$, $B_2$, $I$ and $J$ which satisfy the deformed ADHM equations
\begin{eqnarray}
 &&[B_1,B_1^{\dag}]+[B_2,B_2^{\dag}]+II^{\dag}-J^{\dag}J=2\zeta,
  \label{EQ:ADHM1}\\
 &&[B_1,B_2]+IJ=0,\label{EQ:ADHM2}
\end{eqnarray}
where $B_1$ and $B_2$ are $k\times k$ complex matrices,
$I$ and $J^{\dag}$ are $k\times 1$ complex matrices.
We call this $k$ ``instanton number''.
Note that the right hand side of Eq.(\ref{EQ:ADHM1}) is caused by
the noncommutativity of space.
In $U(1)$ cases, if $\zeta >0$, Eq.(\ref{EQ:ADHM2}) and the stability
condition allow us to take $J=0$ \cite{Nakajima2}\cite{Furuuchi2}.
In \cite{Nekrasov2}\cite{Nekrasov3}, a projector which project the
Hilbert space $\cal H$ to the subspace of $\cal H$ is introduced by
\begin{equation}
 P=I^{\dag}e^{\sum_{\alpha}\beta_{\alpha}^{\dag}\cd_{\alpha}}
  \cket{0}{0}G^{-1}\bra{0}{0}
  e^{\sum_{\alpha}\beta_{\alpha}c_{\alpha}}I,\label{EQ:projection}
\end{equation}
where $B_{\alpha}=\sqrt{\zeta}\beta_{\alpha}$ and
$G$ is a normalization factor (hermitian matrix)
\begin{equation}
 G=\bra{0}{0}e^{\sum_{\alpha}\beta_{\alpha}c_{\alpha}}II^{\dag}
  e^{\sum_{\alpha}\beta_{\alpha}^{\dag}\cd_{\alpha}}\cket{0}{0}.
\end{equation}
Let $S$ be a shift operator which obey the following relations:
\begin{equation}
 SS^{\dag}=1,\;\;S^{\dag}S=1-P.\label{EQ:par-iso}
\end{equation}

Using the shift operator $S$, the $U(1)$ ASD gauge fields
are given as
\begin{equation}
 D_{\alpha}=\sqrt{\frac{1}{\zeta}}S\Lambda^{-\frac{1}{2}}c_{\alpha}
  \Lambda^{\frac{1}{2}}S^{\dag},\;\;
  D_{\alb}=-\sqrt{\frac{1}{\zeta}}S\Lambda^{\frac{1}{2}}\cd_{\alpha}
  \Lambda^{-\frac{1}{2}}S^{\dag},\label{EQ:gauge-NS}
\end{equation}
where
\begin{equation}
 \Lambda=1+I^{\dag}\Deltah^{-1}I,\;\;
  \Deltah=\zeta\sum_{\alpha}(\beta_{\alpha}-\cd_{\alpha})
  (\beta_{\alpha}^{\dag}-c_{\alpha}).\label{EQ:Lambda}
\end{equation}
Note that the subspace projected by $P$ is the kernel of $\hat\Delta$,
and Eq.(\ref{EQ:par-iso}) implies that $\Delta ^{-1}$ is well-defined
since $S^\dag $ (or $S$) removes the kernel.

\section{Instanton charge : the elongated $U(1)$ instanton case}
\label{SEC:elongated}
In \cite{Ishikawa1}, we calculated the instanton charge (integral of the
Pontrjagin class) of elongated $U(1)$ instanton numerically. From the
results,
it is expected that the instanton charge is integer and
equal to instanton number as same as commutative case.
Therefore, we justify that statement analytically.

\subsection{Elongated $U(1)$ instantons on noncommutative $\bf R^4$}
In this subsection, we review the elongated type $U(1)$
instanton solutions with arbitrary instanton number $k$ in the
noncommutative space \cite{Ishikawa1}.
After that, we investigate the property of these solutions to
calculate the instanton charge.
In the following, we set the noncommutativity of space as
$\zeta_1=\zeta_2=\zeta$.\\

We set the ADHM data as
\begin{equation}
 B_1=\sum_{i=1}^{k-1}\sqrt{2i\zeta}e_ie_{i+1}^{\dag},\;\;
  B_2=0,\;\;I=\sqrt{2k\zeta}e_k,\;\;J=0,\label{EQ:B1B2IJ}
\end{equation}
where $e_i$ is defined by
\begin{equation}
 e_i^{\dag}=
 (\stackrel{1}{0},\stackrel{}{\cdots},\stackrel{i-1}{0},\stackrel{i}{1},
 \stackrel{i+1}{0},\stackrel{}{\cdots},\stackrel{k}{0}).
\end{equation}
These matrices (\ref{EQ:B1B2IJ}) satisfy the deformed ADHM equations
(\ref{EQ:ADHM1}) and (\ref{EQ:ADHM2}).
These ADHM data correspond to the configuration
that $k$ instantons are elongated into $z_1$-$\bz_1$ direction.
Above ADHM data give the expression for the projection $P$
(\ref{EQ:projection}) as
\begin{equation}
 P=I^{\dag}e^{\sum_{\alpha}\beta_{\alpha}^{\dag}\cd_{\alpha}}
  \cket{0}{0}G^{-1}\bra{0}{0}
  e^{\sum_{\alpha}\beta_{\alpha}c_{\alpha}}I
  =\sum_{n_1=0}^{k-1}\cket{n_1}{0}\bra{n_1}{0}\;.
  \label{EQ:projection-elongated}
\end{equation}
This $P$ induces the shift operator (\ref{EQ:par-iso}) :
\begin{equation}
 S^{\dag}=\sum_{n_1=0}^{\infty}\cket{n_1+k}{0}\bra{n_1}{0}
  +\sum_{n_1=0}^{\infty}\sum_{n_2=1}^{\infty}\cket{n_1}{n_2}\bra{n_1}{n_2}.
  \label{EQ:elongate-S}
\end{equation}
$\Lambda(\hat{n}_1,\hat{n}_2)$ in gauge fields (\ref{EQ:gauge-NS}) is
obtained as follows :
\begin{equation}
 \Lambda(\hat{n}_1,\hat{n}_2)
  =\frac{w_k(\hat{n}_1,\hat{n}_2)}
  {w_k(\hat{n}_1,\hat{n}_2)-2kw_{k-1}(\hat{n}_1,\hat{n}_2)},
\end{equation}
where $w_i(n_1,n_2)$ is generated by $F(t)$ :
\begin{eqnarray}
 w_i(n_1,n_2)&=&\left.\left(\frac{d}{dt}\right)^iF(t)\right|_{t=0},\\
 F(t)&=&(1-t)^{-n_1+n_2+k-1}(1-2t)^{-n_2-1}
  =\sum_{i=0}^{\infty}\frac{w_i}{i!}t^i.
\end{eqnarray}\\

\begin{figure}[t]
 \begin{center}
   \scalebox{0.5}{\includegraphics{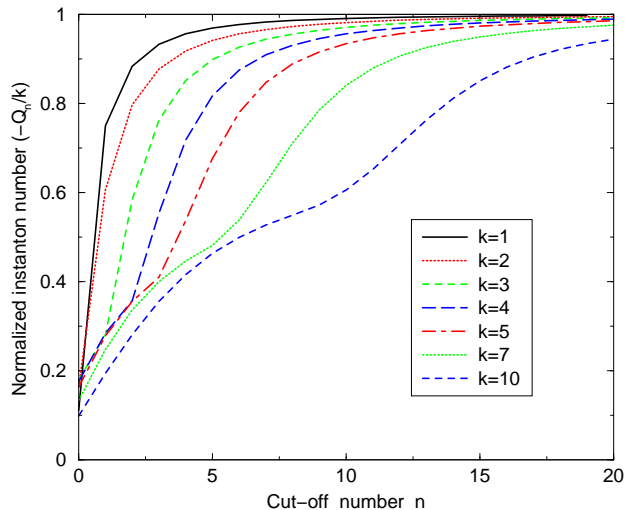}}
 \end{center}
 \caption{Convergency of the instanton charge \cite{Ishikawa1};
 As the cut-off number $n$ increases, all the normalized
 instanton numbers $-Q_n/k$ approach one.
 These results show that the instanton charge is equivalent to the
 instanton number numerically.}
 \label{FIG:conv}
\end{figure}

In the paper \cite{Ishikawa1}, we checked the instanton charge $Q$ of our
solution numerically.
In its computation, we introduced the Fock space cut-off $n$ as
\begin{eqnarray}
 Q&=&\lim_{n\rightarrow\infty}Q_n,\\
 Q_n&=&\zeta^2\sum_{n_1=0}^n\sum_{n_2=0}^n
  \bra{n_1}{n_2}(F_{1\bar{1}}F_{2\bar{2}}-F_{1\bar{2}}F_{2\bar{1}})
  \cket{n_1}{n_2}.
\end{eqnarray}
The result is shown in Fig.\ref{FIG:conv}.
This result is enough to confirm that the Pontrjagin class of this
solution is equivalent to the instanton
number in the limit $n\rightarrow\infty$.

\subsection{Constraints from ASD conditions}
We consider constraints on the instanton solution
(\ref{EQ:gauge-NS}) from anti-self-dual (ASD) conditions. It is needless
to say that the solution (\ref{EQ:gauge-NS}) satisfies the ASD conditions
because it is constructed by the ADHM method. However, it is too difficult
to get the equations (recursion relations) of $\Lambda$
(\ref{EQ:Lambda}) as explicit forms. This fact make difficulty to
analyze the instanton charge analytically. Therefore, we substitute
Eq.(\ref{EQ:gauge-NS}) for ASD conditions directly and obtain the
conditions for $\Lambda$, in this section.

The instanton solution (\ref{EQ:gauge-NS}) should satisfy the ASD
conditions (\ref{EQ:ASD}):
\begin{eqnarray}
 &&[D_1,D_2]=[D_{\bar{1}},D_{\bar{2}}]=0\;,\label{EQ:ASD-D1}\\
 &&[D_1,D_{\bar{1}}]+[D_2,D_{\bar{2}}]=-\frac{2}{\zeta}\;.\label{EQ:ASD-D2}
\end{eqnarray}
We substitute the solution (\ref{EQ:gauge-NS}) for
Eqs.(\ref{EQ:ASD-D1}) and (\ref{EQ:ASD-D2}), then
\begin{eqnarray}
 &&[D_1,D_2]\nonumber\\
 &&=
  \frac{1}{\zeta}S\Bigl\{\Lambda^{-\frac{1}{2}}(\htn_1,\htn_2)c_1
  \Lambda^{\frac{1}{2}}(\htn_1,\htn_2)(1-P)
  \Lambda^{-\frac{1}{2}}(\htn_1,\htn_2)c_2
  \Lambda^{\frac{1}{2}}(\htn_1,\htn_2)
  -(c_1\leftrightarrow c_2)\Bigr\}S^{\dag}\nonumber\\
 &&=0,\label{EQ:constraint1}\\
 &&[D_1,D_{\bar{1}}]+[D_2,D_{\bar{2}}]\nonumber\\
 &&=
  -\frac{1}{\zeta}S\Bigl\{
  \Lambda^{-\frac{1}{2}}(\htn_1,\htn_2)c_1
  \Lambda^{\frac{1}{2}}(\htn_1,\htn_2)(1-P)
  \Lambda^{\frac{1}{2}}(\htn_1,\htn_2)c_1^{\dag}
  \Lambda^{-\frac{1}{2}}(\htn_1,\htn_2)\Bigr.\nonumber\\
 &&\hspace{13mm}
  \Bigl.
  -\Lambda^{\frac{1}{2}}(\htn_1,\htn_2)c_1^{\dag}
  \Lambda^{-\frac{1}{2}}(\htn_1,\htn_2)(1-P)
  \Lambda^{-\frac{1}{2}}(\htn_1,\htn_2)c_1
  \Lambda^{\frac{1}{2}}(\htn_1,\htn_2)+(c_1\leftrightarrow c_2)
  \Bigr\}S^{\dag}\nonumber\\
 &&=
  -\frac{2}{\zeta}.\label{EQ:constraint2}
\end{eqnarray}
$\Lambda^{\frac{1}{2}}(n_1+1,n_2)/\Lambda^{\frac{1}{2}}(n_1,n_2)$ and
$\Lambda^{\frac{1}{2}}(n_1,n_2+1)/\Lambda^{\frac{1}{2}}(n_1,n_2)$ appear
frequently in the following discussion.
Therefore we introduce $h_{\alpha}(n_1,n_2)$ for convenience:
\begin{equation}
 \frac{\Lambda(n_1+1,n_2)}{\Lambda(n_1,n_2)}=1-h_1(n_1,n_2),\hspace{5mm}
  \frac{\Lambda(n_1,n_2+1)}{\Lambda(n_1,n_2)}=1-h_2(n_1,n_2).
  \label{EQ:h-definition}
\end{equation}
The projector $P$ is not generally diagonalized in the number operator
representation, that is, $\langle n_1,~n_2|P|n'_1,~n'_2\rangle$ is not
proportional to $\delta _{n_1 n'_1}\delta _{n_2 n'_2}$ for arbitrary $P$.
However there is a class that  $P$ is diagonalized, which includes
elongated $U(1)$ instantons as we saw in
Eq.(\ref{EQ:projection-elongated}).
General case is discussed in
section \ref{SEC:general}, but its calculation is complex. Therefore in
this section we treat only diagonal $P$ case as a simple example.

In such cases, Eqs.(\ref{EQ:constraint1}) and (\ref{EQ:constraint2}) are
rewritten as
\begin{eqnarray}
 &&h_1(n_1,n_2+1)-h_1(n_1,n_2)-h_2(n_1+1,n_2)+h_2(n_1,n_2)\nonumber\\
 &&\hspace{15mm}-h_1(n_1,n_2+1)h_2(n_1,n_2)+h_1(n_1,n_2)h_2(n_1+1,n_2)=0,
  \label{EQ:h-relation1}\\
 &&(n_1+1)h_1(n_1,n_2)+(n_2+1)h_2(n_1,n_2)\nonumber\\
 &&\hspace{15mm}-n_1h_1(n_1-1,n_2)-n_2h_2(n_1,n_2-1)=0
  \hspace{10mm}\Bigl((n_1,n_2)\not\in{\cal P}\Bigr), \label{EQ:ASD45}
\end{eqnarray}
with
\begin{equation}
 h_{\alpha}(n_1,n_2)=1\hspace{10mm}\Bigl((n_1,n_2)\in{\cal P}\Bigr),
\end{equation}
where $\cal P$ represents the subspace of Fock space which is projected
out by $P$.
These recursion relations are used when we estimate
the instanton charge in the next subsection.

\subsection{Instanton number : elongated instanton case}
\label{SEC:elongated-number}
In this subsection, we calculate analytically
the instanton charge of elongated $U(1)$ instantons without the explicit
form of the solution and show the charge is equal to the instanton number.
Elongated $U(1)$ instantons belong to the class of the instantons
whose projection operator $P$ is diagonalized in the number operator
representation. This is a good exercise to analyze a general solution
in section \ref{SEC:general}.

\begin{figure}[t]
  \begin{center}
   \begin{minipage}{70mm}
    \begin{center}
     \psfragscanon
     \psfrag{n1}[][][1.5]{$n_1$}
     \psfrag{n2}[][][1.5]{$n_2$}
     \psfrag{0} [][][1.5]{$0$}
     \psfrag{p1}[][][1.5]{$k-1$}
     \psfrag{p2}[][][1.5]{1}
     \psfrag{c1}[][][1.5]{$N$}
     \psfrag{c2}[][][1.5]{$N$}
     \psfrag{P} [][][1.5]{$P$}
     \psfrag{R} [][][2.0]{$R_0$}
     \scalebox{0.5}{\includegraphics{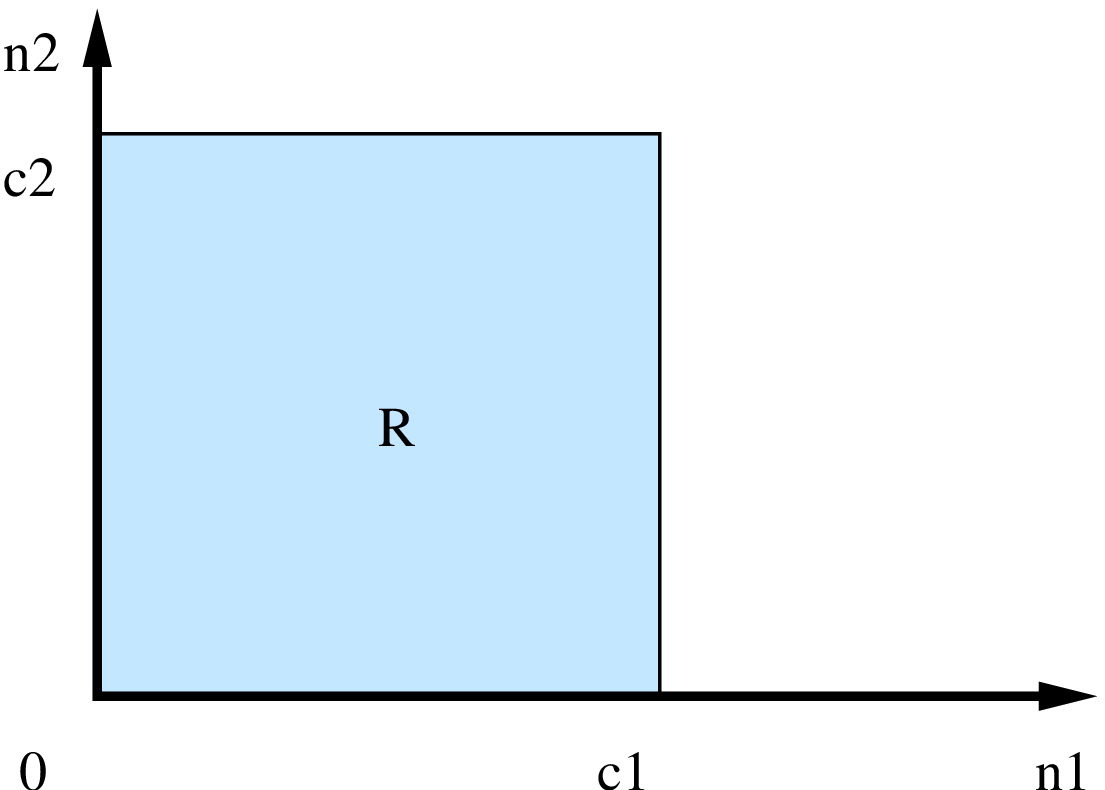}}
    \end{center}
    \caption{Summation rule (region of $\mbox{Tr}_{\cal H}$ operation);
    This represents the region of summation of $\mbox{Tr}_{\cal H}$
    operation.
    We set the cut-off of Fock space in the $n_1$ and $n_2$ direction at
$N$.}
    \label{FIG:region0}
   \end{minipage}
   \hspace{5mm}
   \begin{minipage}{70mm}
    \begin{center}
     \psfragscanon
     \psfrag{n1}[][][1.5]{$n_1$}
     \psfrag{n2}[][][1.5]{$n_2$}
     \psfrag{0} [][][1.5]{$0$}
     \psfrag{p1}[][][1.5]{$k-1$}
     \psfrag{p2}[][][1.5]{$0$}
     \psfrag{c1}[][][1.5]{$N$}
     \psfrag{c2}[][][1.5]{$N$}
     \psfrag{P} [][][1.5]{$\cal P$}
     \psfrag{R} [][][2.0]{$R_S$}
     \psfrag{s1}[][][1.5]{$N+k$}
     \scalebox{0.5}{\includegraphics{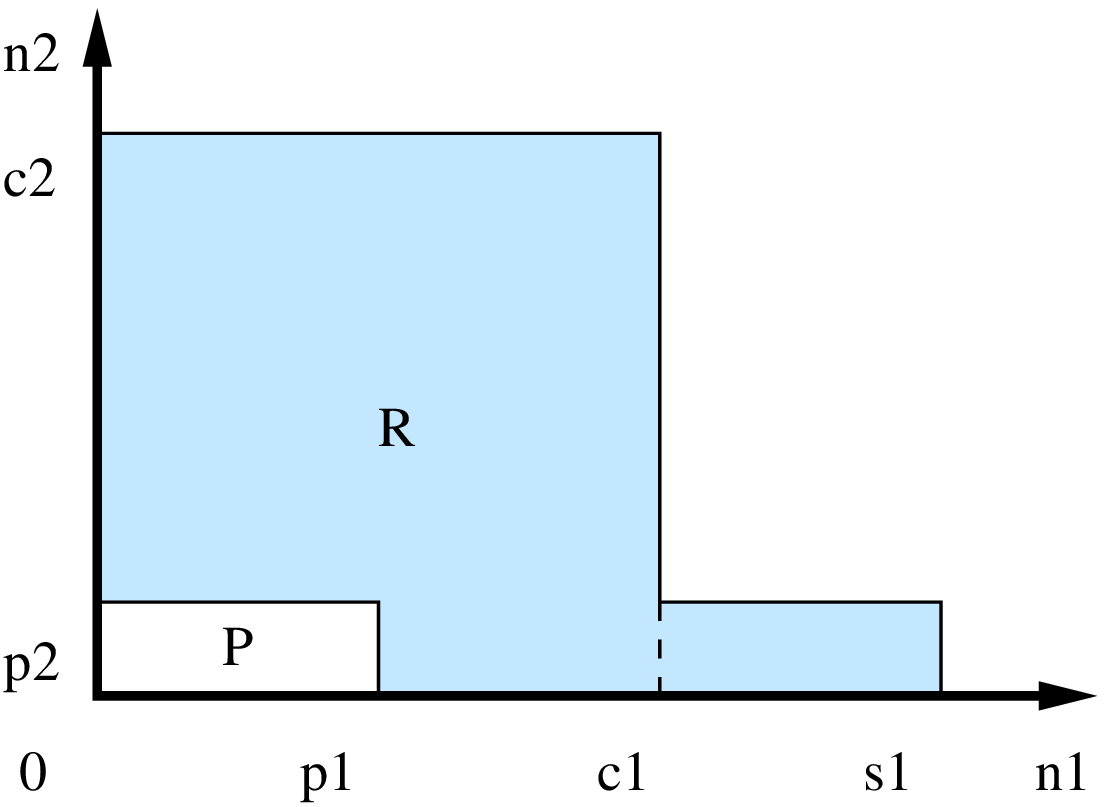}}
    \end{center}
    \caption{Summation rule (inside of $S$ and $S^{\dag}$);
    This represents the region of intermediate summation of Fock space
    that is carried out inside of $S$ and $S^{\dag}$.\vspace{7mm}}
    \label{FIG:regionS}
   \end{minipage}
  \end{center}
\end{figure}

The key point is to introduce the cut-off to calculate
the instanton charge.
We set the cut-off in the $n_1$ and $n_2$ directions on $N$ ($N\gg k$)
(Fig.\ref{FIG:region0}),
here the ``cut-off'' means the trace operation cut-off, that is,
\begin{equation}
 {\rm Tr}_{\cal H}\left|_{\scsc [0,~N]}
     O\right.
 =
 \sum_{n_1=0}^{N}\sum_{n_2=0}^{N}
 \bra{n_1}{n_2}O\cket{n_1}{n_2} \;,
\end{equation}
where $O$ is an arbitrary operator.
Note that this cut off is only for the upper bound of summation of the
initial state and the final state. To the contrary, upper bound of
intermediate summation is frequently over $N$. To see this phenomena let
us consider the trace of Eq.(\ref{EQ:ASD-D2}).
The trace operation of Eq.(\ref{EQ:ASD-D2}) is typically of the form:
\begin{equation}
 \mbox{Tr}_{\cal H}\big|_{\scsc [0,~N]}
\bigl(SO'S^{\dag}\bigr)
  =\sum_{n_1=0}^{N}\sum_{n_2=0}^{N}
  \bra{n_1}{n_2}SO'S^{\dag}\cket{n_1}{n_2}\;,
\end{equation}
where $O'$ is an arbitrary operator,
so summing up the Fock space about $O'$ (sandwiched between $S$ and
$S^{\dag}$) actually include the outside of cut-off.
For example, in the case of elongated instantons (\ref{EQ:elongate-S})
implies that the above expression is rewritten as
\begin{eqnarray}
 \mbox{Tr}_{\cal H}\big|_{\scsc [0,~N]}
 \bigl(SO'S^{\dag}\bigr)&=&
  \sum_{n_1=k}^{N+k}\bra{n_1}{0}O'\cket{n_1}{0}
  +\sum_{n_1=0}^{N}\sum_{n_2=1}^{N}
  \bra{n_1}{n_2}O'\cket{n_1}{n_2}\nonumber\\
 &=&\sum_{n_1=0}^{\infty}\sum_{n_2=0}^{\infty}
  \bra{n_1}{n_2}O'\cket{n_1}{n_2}\theta((n_1,n_2)\in R_s),
\end{eqnarray}
where $\theta(X)$ is the $\theta$-function whose value is $1$ if the
proposition $X$ is true and $0$ if $X$ is false (Fig.\ref{FIG:regionS}).\\

For later convenience, we take a trace of ASD condition (\ref{EQ:ASD45}) :
\begin{eqnarray}
 &&\mbox{Tr}_{\cal H}\big|_{\scsc [0,~N]}
    \Bigl\{
  2+\zeta[D_1,D_{\bar{1}}]+\zeta[D_2,D_{\bar{2}}]\Bigr\}\nonumber\\
 &&=\sum_{n_1=0}^{\infty}\sum_{n_2=0}^{\infty}
  \Bigl\{(n_1+1)h_1(n_1,n_2)+(n_2+1)h_2(n_1,n_2)\Bigr.\nonumber\\
 &&\hspace{20mm}\Bigl.-n_1h_1(n_1-1,n_2)-n_2h_2(n_1,n_2-1)\Bigr\}
  \theta((n_1,n_2)\in R_s)\nonumber\\
 &&=(N+k+1)h_1(N+k,0)-k+(N+1)\sum_{n_2=1}^{N}h_1(N,n_2)\nonumber\\
 &&\hspace{15mm}
  +(N+1)\sum_{n_1=0}^{N}h_2(n_1,N)-k+\sum_{n_1=N+1}^{N+k}h_2(n_1,0)
  \nonumber\\
 &&=0.\label{EQ:sum-const2}
\end{eqnarray}
Note that in the right hand side of the first equality,
only progression of differences
appear, so the summation on the cut-off remains. This fact will be
generalized as ``Stokes' theorem`` in the next section. From the above
result we obtain the relation
\begin{eqnarray}
 E&\equiv&(N+1)\sum_{n_1=0}^{N}h_2(n_1,N)
  +(N+1)\sum_{n_2=1}^{N}h_1(N,n_2)\nonumber\\
 &&+\sum_{n_1=N+1}^{N+k}h_2(n_1,0)+(N+k+1)h_1(N+k,0)\nonumber\\
 &=&2k\;.\label{EQ:E-equation}
\end{eqnarray}

Now let us calculate the instanton charge $Q$. $Q$ is possible to be
written with progression of differences.
The detail of the calculation is presented in appendix
\ref{AP:charge-elongated}.
The result is
\begin{equation}
 Q_{\rm cutoff}=k-E+(h_{\alpha}^2 \;\mbox{terms})
  =-k+(h_{\alpha}^2 \;\mbox{terms}),
\end{equation}
where $Q_{\rm cutoff}$ means
\begin{eqnarray}
 Q_{\rm cutoff}&=&
  \zeta^2\mbox{Tr}_{\cal H}\big|_{\scsc [0,~N]}
  \bigl(F_{1\bar{1}}F_{2\bar{2}}-F_{1\bar{2}}F_{2\bar{1}}),\\
 Q&=&\lim_{N\rightarrow\infty}Q_{\rm cutoff}.
\end{eqnarray}
Therefore, except for the $h_{\alpha}^2$ terms, we obtain the instanton
charge $-k$.
To identify the charge as the instanton number, we need to take
the limit $N\rightarrow\infty$.
Then we have to know that the behavior of $\hat{\Delta}$ in
Eq.(\ref{EQ:Lambda}) for the Fock state $|n_1,~n_2\rangle$ in the
limit $n_1\;\mbox{or}\;n_2\rightarrow\infty$.
By definition Eq.(\ref{EQ:Lambda} ) $\hat{\Delta}$ behaves in this limit as
\begin{equation}
 \hat{\Delta}\rightarrow\zeta(\hat{n}_1+\hat{n}_2)\hspace{10mm}
  (n_{\alpha}\rightarrow\infty)\;.
\end{equation}
 From Eq.(\ref{EQ:Lambda})
 thus $\Lambda(\hat{n}_1,\hat{n}_2)$  is obtained as
\begin{equation}
 \Lambda(\hat{n}_1,\hat{n}_2)\rightarrow 1+\frac{1}{\hat{n}_1+\hat{n}_2}
  \hspace{10mm}(n_{\alpha}\rightarrow\infty)\;,\label{EQ:Lambda-behavior}
\end{equation}
which implies
\begin{equation}
    \frac{\Lambda(\hat{n}_1+1,\hat{n}_2)}
          {\Lambda(\hat{n}_1,\hat{n}_2)}\;\;
     {\rm and} \;\;
          \frac{\Lambda(\hat{n}_1,\hat{n}_2+1)}
             {\Lambda(\hat{n}_1,\hat{n}_2)}
               \ \rightarrow \
              1-\frac{1}{(\hat{n}_1+\hat{n}_2+1)^2}
                 \hspace{7mm}
          (n_{\alpha}\rightarrow\infty)\;.
\end{equation}
Therefore, $h_{\alpha}(\htn_1,\htn_2)$ approaches to
\begin{equation}
 h_{\alpha}(\htn_1,\htn_2)\rightarrow\frac{1}{(\hat{n}_1+\hat{n}_2)^2}
  \hspace{10mm}(n_{\alpha}\rightarrow\infty)\;.
\end{equation}
As a result, the $h_{\alpha}^2$ terms vanish in the limit
$N\rightarrow\infty$.
Finally, we obtain the instanton number:
\begin{equation}
 Q=\lim_{N\rightarrow\infty}Q_{\rm cutoff}=-k\;.
\end{equation}

Above calculation is performed only in the elongated $U(1)$ instanton case.
However, as long as the projection operator $P$ is diagonalized in the
number operator representation,
the calculation for other type instantons is same as above.
In such case if we put $N$ sufficiently large, then $k$ cells
(Fock states) are removed from the $(N+1)\times (N+1)$ square
(Fig.\ref{FIG:regionS}) and extra $k$ cells are added outside of the square.
Their added $k$ cells' location is arbitrary chosen.
However it is easy to show that $Q$ do not depend on the location.

Contrary for the case including the projector $P$ cannot be diagonalized,
the calculation is altered.
The next section contains to study such cases.

\section{Instanton charge : general case}
\label{SEC:general}
In this section, we construct the instanton charge for
general $U(1)$ instanton case.
The instanton charge is defined by an integral of the Pontrjagin class
(\ref{EQ:Q-op-N}). We find out that it is possible to define the instanton
charge under some converge condition, especially the charge
 looks like conditionally
converge. Finally, we see that the instanton charge is equivalent to
instanton
number defined in ADHM construction, as long as the converge condition
is satisfied.
\subsection{Instanton equations }
In the previous section,
the projector $P$ is restricted to the special form of
$\sum\cket{n_1}{n_2}\bra{n_1}{n_2} $.
In that case, instanton charge was defined clearly
and calculation with the cut-off was easily done.
In this section, we consider the instanton number of
an arbitrary instanton solution.
The general solution of the instanton is formally given by Nekrasov
in \cite{Nekrasov2}\cite{Nekrasov3} as Eq.(\ref{EQ:gauge-NS}).
This solution is given formally by ADHM construction.
Therefore $\Lambda^{\frac{1}{2}}(\hat{n}_1,\hat{n}_2)$ , $P$ and $S$ are
possible to be determined by ADHM data formally.
However, we do not show concrete expression of them in practice, since
the calculation for multi-instantons is too complex.
To avoid this problem, though we give the form of instanton solution
(\ref{EQ:gauge-NS})
as a result of ADHM construction, we also find constraints condition of
$\Lambda^{\frac{1}{2}}(\hat{n}_1,\hat{n}_2)$ , $P$ and $S$ not by ADHM
data but directly ASD equations (\ref{EQ:ASD-D1}) and (\ref{EQ:ASD-D2}).

At first, let us make a more concrete form of $S$.
For example, the eigenvectors of the projector (\ref{EQ:projection}) are
expressed as
\begin{eqnarray}
a^i \equiv I^{\dagger}
 e^{\sum_{\alpha} \beta^{\dagger}_{\alpha} c^{\dagger}_{\alpha}}
 \cket{0}{0} e_i\hspace{10mm}(i=1 \sim k), \label{g-3}
\end{eqnarray}
where $e_i$ is a base of $k$ dimensional vector.
$P$ is a projector onto the $k$ dimensional subspace and
$S^{\dagger}$ is a map to orthogonal complement of
this subspace.
We denote the basis of the subspace as $a^i$.
Note that some model might exist such that this expression (\ref{g-3})
is not valid to make $k$ independent vector. But the following study
does not depend on the concrete expression of $a^i$.
We denote the bases of the orthogonal complement as $a^j_{\bot} (j \ge
k+1)$:
\begin{eqnarray}
(a^j_{\bot})^{\dagger} a^i=0\ .
\end{eqnarray}
In the case of (\ref{g-3}), this condition is written as:
\begin{eqnarray}
     (a^j_{\bot})^{\dagger} a^i
  &=&
        \overline{(a^j_{\bot})}_{n_1 n_2}
         \bra{n_1}{n_2}
       I^{\dagger}e^{\sum_{\alpha}
           \beta^{\dagger}_{\alpha}
         c^{\dagger}_{\alpha}}
             \cket{0}{0} e_i
              \nonumber \\
  &=&
        \overline{(a^j_{\bot})}_{n_1 n_2}
         \bra{0}{0}
          \frac{c^{n_1}_1}{\sqrt{n_1 !}}
           \frac{c^{n_2}_2}{\sqrt{n_2 !}}
         I^{\dagger}e^{\sum_{\alpha}
             \beta^{\dagger}_{\alpha}
           c^{\dagger}_{\alpha}}
               \cket{0}{0} e_i \\
  &=&
         \overline{(a^j_{\bot})}_{n_1 n_2}
       I^{\dagger}\frac{\beta^{\dag n_1}_1}{\sqrt{n_1 !}}
           \frac{\beta^{\dag n_2}_2}{\sqrt{n_2 !}}e_i =0,
            \nonumber
\end{eqnarray}
where we expand $a^j_{\bot}$ by the Fock space $\bra{n_1}{n_2}$ with its
coefficient $(a^j_{\bot})_{n_1 n_2}$ and $\overline{(a^j_{\bot})}_{n_1 n_2}$
is a complex conjugate of $(a^j_{\bot})_{n_1 n_2}$.
We choose these basis to satisfy the orthonormal conditions,
\begin{eqnarray}
\sum_{n_1, n_2}^{\infty } \overline{(a^i_{\bot})}_{n_1 n_2}
(a^j_{\bot})_{n_1 n_2}= \delta^{ij}.
\end{eqnarray}
The dimension of the $a^j_{\bot}$ is symbolically $\infty - k$ and
$\{a^i\} \bigoplus \{ a^j_{\bot}\}$ is a complete system.
For convenience, we rewrite the basis as
\begin{eqnarray}
A^i=\begin{cases}
     a^i &(i=1,2,\cdots k)\\
     a^i_{\bot} &(i=k+1,\cdots,\infty)\\
    \end{cases} .\label{g-7}
\end{eqnarray}
With this $A^i$ we can express $S$ and $S^{\dagger}$ concretely as
\begin{eqnarray}
      S^{\dagger}
   &=&
          \sum_{n,m,n',m',i,j}
           \overline{A}^{i+k}_{n',m'}
         S^{\dagger}_{ij}
             \cket{n'}{m'}
              \bra{n}{m}A^j_{n,m},
               \nonumber\\
      S
   &=&
         \sum_{n,m,n',m',i,j}
          \overline{A}^{i}_{n',m'}
     S_{ij}\cket{n'}{m'}
            \bra{n}{m}A^{j+k}_{n,m}.
             \label{EQ:shift-general}
\end{eqnarray}
Since these operators have to obey the relation $SS^{\dagger}=1$, the
coefficients $A^i_{n,m}$ have to satisfy the normalization condition:
\begin{eqnarray}
 \sum_{i,j,p=0}^{\infty}\overline{A}^i_{n,m}S_{ij}
  S^{\dagger}_{jp}A^p_{n',m'}
  =\delta_{nn'}\delta_{mm'}.\label{g-8}
\end{eqnarray}
Additionally $S^{\dagger}S=1-P$ is expressed as the following form,
\begin{eqnarray}
 1-P&=&\sum_{{\rm all\ indices}=0}^{\infty} 
  \overline{A}^{i+k}_{n,m}S^{\dagger}_{ij}A^j_{n',m'}
  \overline{A}^{l}_{n',m'}S_{lp}A^{p+k}_{n'',m''}
  \cket{n}{m}\bra{n''}{m''}\\
 &=& \sum_{{\rm all\ indices}=0}^{\infty} 
  \overline{A}^{i+k}_{n,m}S^{\dagger}_{ij}
  S_{jp}A^{p+k}_{n',m'}
  \cket{n}{m}\bra{n'}{m'}.\label{g-9}
\end{eqnarray}
Note that the operators $S$ and $S^\dag$ are partial isometry operators,
but we can
choose the matrices $S_{ij}$ and $S_{ij}^\dag$ as unitary matrices, since
the shift of upper indices $i$ of $A^i_{nm}$ in
Eq.(\ref{EQ:shift-general}) is possible to play a role of a shift operator.
For convenience, we introduce $K_{n,m,n',m'}$ :
\begin{equation}
 K_{n,m,n',m'}=\sum\overline{A}^{i+k}_{n,m}S^{\dagger}_{ij}
  S_{jp}A^{p+k}_{n',m'}.
\end{equation}
Since $(1-P)$ is a projector i.e. $(1-P)^2= 1-P$ ,
then $K_{n,m,n',m'}$ have to satisfy the following condition:
\begin{eqnarray}
 \sum_{n',m'} K_{n,m,n',m'}K_{n',m',l,p}=K_{n,m,l,p}. \label{g-9.1}
\end{eqnarray}\\

The next step, let us make a concrete $\Lambda$ expression of ASD
Eqs.(\ref{EQ:constraint1}) and (\ref{EQ:constraint2}).
We substitute Eqs.(\ref{EQ:gauge-NS}) and (\ref{g-9}) for
Eq.(\ref{EQ:ASD-D1}) as
\begin{eqnarray}
 &&[D_1,D_2]\nonumber\\
 &&\hspace{-5mm}=\frac{1}{\zeta}S\Bigl\{
  \sum
  K_{n',m',n,m}
  \frac{\Lambda^{\frac{1}{2}}(n',m')}{\Lambda^{\frac{1}{2}}(n'-1,m')}
  \frac{\Lambda^{\frac{1}{2}}(n,m+1)}{\Lambda^{\frac{1}{2}}(n,m)}
  \sqrt{n'(m+1)}  \cket{n'-1}{m'} \bra{n}{m+1}
  \nonumber \\
 &&\hspace{3mm}-\sum 
  K_{n',m',n,m}
  \frac{\Lambda^{\frac{1}{2}}(n',m')}{\Lambda^{\frac{1}{2}}(n',m'-1)}
  \frac{\Lambda^{\frac{1}{2}}(n+1,m)}{\Lambda^{\frac{1}{2}}(n,m)}
  \sqrt{m'(n+1)} \cket{n'}{m'-1} \bra{n+1}{m}
  \Bigl\}S^{\dag}\nonumber \\
 &&\hspace{-5mm}=0.\label{g-10}
\end{eqnarray}
Note that we denote $\htn$ as a number operator and
$n$ as a C-number.
 From Eq.(\ref{g-10}), instanton equation Eq.(\ref{EQ:ASD-D1})
is rewritten as
\begin{eqnarray}
 &&K_{n'+1,m',n,m-1}\sqrt{(n'+1)m}
  \frac{\Lambda^{\frac{1}{2}}(n'+1,m')}{\Lambda^{\frac{1}{2}}(n,m-1)}
  \nonumber\\
 &&\hspace{20mm}=
  K_{n',m'+1,n-1,m}\sqrt{(m'+1)n}
  \frac{\Lambda^{\frac{1}{2}}(n',m'+1)}{\Lambda^{\frac{1}{2}}(n-1,m)}.
  \label{g-11}
\end{eqnarray}

 From Eq.(\ref{EQ:ASD-D2}), we get the another equation:
\begin{eqnarray}
 &&[D_1,D_{\bar{1}}]+[D_2,D_{\bar{2}}]\nonumber\\
 &&=-\frac{1}{\zeta}S
  \left\{
   K_{n',m',n,m}\sqrt{n'n}
   \frac{\Lambda^{\frac{1}{2}}(n',m')}{\Lambda^{\frac{1}{2}}(n'-1,m')}
   \frac{\Lambda^{\frac{1}{2}}(n,m)}{\Lambda^{\frac{1}{2}}(n-1,m)}
   \cket{n'-1}{m'}\bra{n-1}{m} \right. \nonumber \\
 &&-K_{n',m',n,m}\sqrt{(n'+1)(n+1)}
  \frac{\Lambda^{\frac{1}{2}}(n'+1,m')}{\Lambda^{\frac{1}{2}}(n',m')}
  \frac{\Lambda^{\frac{1}{2}}(n+1,m)}{\Lambda^{\frac{1}{2}}(n,m)}
  \cket{n'+1}{m'}\bra{n+1}{m} \nonumber \\
 &&+K_{n',m',n,m}\sqrt{m'm}
  \frac{\Lambda^{\frac{1}{2}}(n',m')}{\Lambda^{\frac{1}{2}}(n',m'-1)}
  \frac{\Lambda^{\frac{1}{2}}(n,m)}{\Lambda^{\frac{1}{2}}(n,m-1)}
  \cket{n'}{m'-1}\bra{n}{m-1}\nonumber \\
 &&\left.-K_{n',m',n,m}\sqrt{(m'+1)(m+1)}
    \frac{\Lambda^{\frac{1}{2}}(n',m'+1)}{\Lambda^{\frac{1}{2}}(n',m')}
    \frac{\Lambda^{\frac{1}{2}}(n,m+1)}{\Lambda^{\frac{1}{2}}(n,m)}
    \cket{n'}{m'+1}\bra{n}{m+1}
   \right\} S^{\dagger}\nonumber\\
 &&=-\frac{2}{\zeta}.\label{g-12}
\end{eqnarray}
 From this equation, another recursion relation is obtained as
\begin{eqnarray}
 2K_{l',p',l,p}&=&K_{l',p',n',m'}K_{n,m,l,p}\nonumber\\
 &&\hspace{-5mm}\times
  \left\{
   K_{n'+1,m',n+1,m}\sqrt{(n'+1)(n+1)}
   \frac{\Lambda^{\frac{1}{2}}(n'+1,m')}{\Lambda^{\frac{1}{2}}(n',m')}
   \frac{\Lambda^{\frac{1}{2}}(n+1,m)}{\Lambda^{\frac{1}{2}}(n,m)}
  \right.\nonumber\\
 &&-K_{n'-1,m',n-1,m}\sqrt{n'n}
  \frac{\Lambda^{\frac{1}{2}}(n',m')}{\Lambda^{\frac{1}{2}}(n'-1,m')}
  \frac{\Lambda^{\frac{1}{2}}(n,m)}{\Lambda^{\frac{1}{2}}(n-1,m)}
  \nonumber\\
 &&+K_{n',m'+1,n,m+1}\sqrt{(m'+1)(m+1)}
  \frac{\Lambda^{\frac{1}{2}}(n',m'+1)}{\Lambda^{\frac{1}{2}}(n',m')}
  \frac{\Lambda^{\frac{1}{2}}(n,m+1)}{\Lambda^{\frac{1}{2}}(n,m)}
  \nonumber\\
 &&\left.
    -K_{n',m'-1,n,m-1}\sqrt{m'm}
    \frac{\Lambda^{\frac{1}{2}}(n',m')}{\Lambda^{\frac{1}{2}}(n',m'-1)}
    \frac{\Lambda^{\frac{1}{2}}(n,m)}{\Lambda^{\frac{1}{2}}(n,m-1)}
    \right\} \ .\label{g-13}
\end{eqnarray}
This equation is equivalent to ASD equation (\ref{EQ:ASD-D2}).

\subsection{Stokes' theorem}
\begin{figure}[t]
  \begin{center}
   \psfragscanon
   \psfrag{N1}[][][1.0]{$N'$}
   \psfrag{N2}[][][1.0]{$N$}
   \psfrag{-}[][][1.2]{$-$}
   \psfrag{=}[][][1.4]{ $=$}
  \psfrag{C1}[][][1.5]{$:~O_{\scsc n_1+1,n_1}$}
  \psfrag{C2}[][][1.5]{$:-O_{\scsc n_1,n_1-1}$}
   \scalebox{.55}{\includegraphics{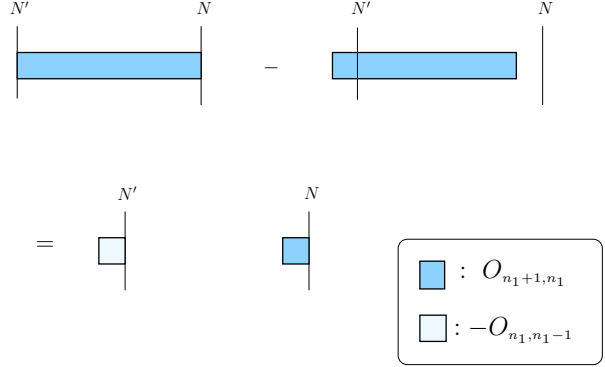}}
  \end{center}
 \caption{Two dimensional Stokes' theorem; integral region is replaced
 by its boundary.}
 \label{FIG:Stokes1}
\end{figure}

\begin{figure}[t]
 \begin{center}
  \psfragscanon
  \psfrag{-}[][][1.5]{$-$}
  \psfrag{+}[][][1.5]{$+$}
  \psfrag{=}[][][1.5]{$=$}
  \psfrag{PLU}[][][1]{$+$}
  \psfrag{C1}[][][1.2]{$:~O_{\scsc n_1+1,n_2,n_1,n_2}$}
  \psfrag{C2}[][][1.2]{$:-O_{\scsc n_1,n_2,n_1-1,n_2}$}
  \psfrag{C3}[][][1.2]{$:~O_{\scsc n_1,n_2+1,n_1,n_2}$}
  \psfrag{C4}[][][1.2]{$:-O_{\scsc n_1,n_2,n_1,n_2-1}$}
  \psfrag{C5}[][][1.2]{$:$}
\scalebox{.73}{\includegraphics{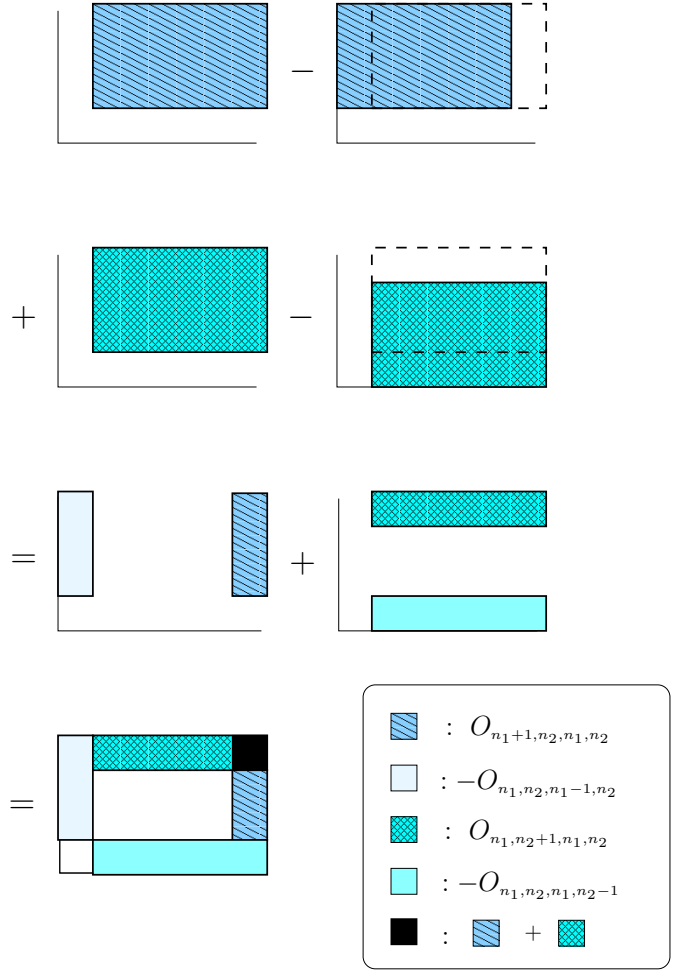}}
 \end{center}
 \caption{Four dimensional case;
 The integral region is a simple rectangle.
 The summation region is replaced by its boundary.}
 \label{FIG:Stokes2}
\end{figure}

\begin{figure}[t]
 \begin{center}
  \psfragscanon
  \psfrag{-}[][][1.5]{$-$}
  \psfrag{+}[][][1.5]{$+$}
  \psfrag{=}[][][1.5]{$=$}
  \psfrag{PLU}[][][1]{$+$}
  \psfrag{C1}[][][1.2]{$:~O_{\scsc n_1+1,n_2,n_1,n_2}$}
  \psfrag{C2}[][][1.2]{$:-O_{\scsc n_1,n_2,n_1-1,n_2}$}
  \psfrag{C3}[][][1.2]{$:~O_{\scsc n_1,n_2+1,n_1,n_2}$}
  \psfrag{C4}[][][1.2]{$:-O_{\scsc n_1,n_2,n_1,n_2-1}$}
  \psfrag{C5}[][][1.2]{$:$}
  \psfrag{C6}[][][1.2]{$:$}
 \scalebox{.73}{\includegraphics{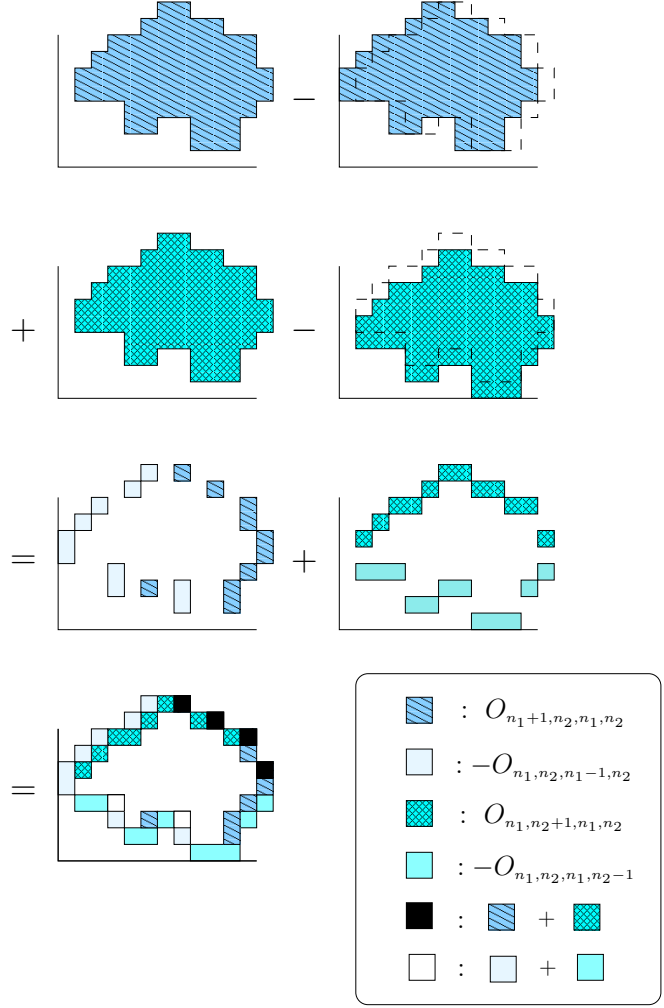}}
 \end{center}
 \caption{Four dimensional case;
 The integral region is an arbitrary case.
 The summation region is replaced by its boundary.}
 \label{intro1}
\end{figure}

In this subsection, we discuss Stokes' like theorem in operator
 representation. Strictly speaking, this is not equivalent to the Stokes'
theorem in usual meaning of commutative space. But we call it simply
``Stokes'
theorem'' in the following. The theorem appears in the calculation
of instanton charge and it is very useful. At first, we consider two
dimensional model:
\begin{eqnarray}
 [c,c^{\dagger}]=1,\hspace{5mm}\htn=c^{\dagger}c.
\end{eqnarray}
Let $O$ be a arbitrary operator, whose Fock state representation is
\begin{eqnarray}
 O=\sum_{n,m}O_{n,m}\left|n\right>\left<m\right|.
\end{eqnarray}
The differentiation by $c^{\dagger}$ is
\begin{eqnarray}
 [c,O]&=&\sum_{n,m}O_{n,m}c\left|n\right>\left<m\right|
  -\sum_{n,m} O_{n,m} \left| n \right> \left< m \right|c \nonumber \\
 &=&\sum_{n,m}O_{n,m}\sqrt{n}\left|n-1\right>\left< m\right|
  -\sum_{n,m}O_{n,m}\sqrt{m+1}\left|n\right>\left<m+1\right|.
  \label{g-19}
\end{eqnarray}
We define the definite integral of the operator $O$
as the trace of the Fock space:
\begin{eqnarray}
 \mbox{Tr}_{\cal H} |_{\scsc [N',N]} O \equiv
  \sum_{n=N'}^{N}  O_{n,n}, \nonumber \\
 \lim_{ N \rightarrow \infty}\mbox{Tr}_{\cal H} |_{\scsc [0,N]}
  = \mbox{Tr}_{\cal H} .
\end{eqnarray}
Then the integral of $[c,O]$ is obtained as
\begin{eqnarray}
 \mbox{Tr}_{\cal H}|_{\scsc [N',N]} [c,O] &=&
  \sum_{n=N'}^N \left( O_{n+1,n} \sqrt{n+1} - O_{n,n-1} \sqrt{n} \right)
  \nonumber\\
 &=& O_{N+1,N} \sqrt{N+1}- O_{N',N'-1} \sqrt{N'}. \label{EQ:82}
\end{eqnarray}
This is the simplest case of the Stokes' theorem
for number operator representation (Fig.\ref{FIG:Stokes1}).
Similarly, the trace of $[c^\dag ~O]$ is
\begin{eqnarray}
\mbox{Tr}_{\cal H}|_{[N',N]} [c^{\dagger},O] =
 O_{N'-1,N'}\sqrt{N'} -O_{N,N+1}\sqrt{N+1}.
\end{eqnarray}
The integral value is determined by the boundary values $N$ and $N'$.
\\
This formula is easily extended to higher dimension.
For example, we consider a four dimensional case.
Let $O$ be an arbitrary operator:
$O=\cket{n}{m}O_{nm,n'm'}\bra{n'}{m'}$.
Then the total derivative of $O$ in the Fock space is
$[c_1,O]+[c_2 ,O]$ and its integration is written as
\begin{eqnarray}
 &&\mbox{Tr}_{\cal H}|_{\rm domain}([c_1, O] + [c_2 , O])\nonumber\\
 &&=\sum_{((n_1,n_2)\in {\rm domain})}
  \bra{n_1}{n_2}([c_1,O]+[c_2,O])\cket{n_1}{n_1}\nonumber\\
 &&={\hspace{-10mm}}\sum_{((n_1,n_2)\in  {\rm boundary})}
  {\hspace{-10mm}}
  \left\{
   O_{n_1+1,n_2,n_1,n_2} \sqrt{n_1+1}
   \theta({\sc n_1+1 \not\in\ {\rm domain}})
   -O_{n_1,n_2,n_1-1,n_2}\sqrt{n_1}
   \theta({\sc n_1-1 \not\in\ {\rm domain}})
  \right. \nonumber \\
 &&\hspace{10mm}\left.
   +O_{n_1,n_2+1,n_1,n_2} \sqrt{n_2+1}
   \theta({\sc n_2+1 \not\in\ {\rm domain}})
   -O_{n_1,n_2,n_1,n_2-1}\sqrt{n_2}
   \theta({\sc n_2-1 \not\in\ {\rm domain}})
    \right\}\nonumber.\\
\label{EQ:stokes-star}
\end{eqnarray}
Here the domain and the boundary are a set of cells belonging some
Fock state and a set of its boundary cells (see Fig.\ref{FIG:Stokes2} and
Fig.\ref{intro1}).
$\theta({\sc n_1+1 \not\in\ {\rm domain}}) $
plays a role of a support on the right side boundary and other
$\theta$ play similar roles.
More higher dimensional case have a similar expression.
On the left side
boundary(${\sc n_1-1 \not\in\ {\rm domain}}$)
and the down side boundary
($ {\sc n_2-1 \not\in\ {\rm domain}}$),
the summation contribute to the trace with minus sign.
 This sign is interpreted as orientation of integral. If
we assign the orientation to the trace, Eq.(\ref{EQ:stokes-star}) is
interpreted as a cyclic integral as same as usual Stokes' theorem.

Next step, let us construct Stokes' theorem
for the covariant derivative with instanton connections
\begin{equation}
 D_{\bar{\alpha}}=-\sqrt{\frac{1}{\zeta}}
  S\Lambda^{\frac{1}{2}}(\htn_1,\htn_2)c_{\alpha}^{\dag}
  \Lambda^{-\frac{1}{2}}(\htn_1,\htn_2) S^{\dag}.
\end{equation}
Since the creation operator is
sandwiched between $S\Lambda^{\frac{1}{2}}(\htn_1,\htn_2)$ and
$\Lambda^{-\frac{1}{2}}(\htn_1,\htn_2)S^{\dag}$,
$[D_{\bar{\alpha}},O]$ is not simple difference form like
Eq.(\ref{g-19}).
However, when we take a trace (integration) of the commutator,
we can find that it is sum of progression of differences :
\begin{eqnarray}
&&\mbox{Tr}_{\cal H}|_{\rm domain}
 [D_{\bar{1}},O]=
 \makebox{}\hspace{-10mm}
 \sum_{n,m,n',m',i,j,i',j'}^{\infty} \hspace{-8mm}
 \overline{A}^{i'}_{n',m'}S_{i'j'}S^{\dagger}_{ij}
 A^j_{n,m}O_{n,m,n',m'}  \label{g-23.5}\\
 && \times \left( \sum_{l,p} A^{j'+k}_{l+1,p}
     \frac{\Lambda^{\frac{1}{2}}(l+1,p)}{ \Lambda^{\frac{1}{2}}(l,p)}
     \overline{A}^{i+k}_{l,p}\sqrt{l+1}
     -A^{j'+k}_{l,p}
     \frac{\Lambda^{\frac{1}{2}}(l,p)}{ \Lambda^{\frac{1}{2}}(l-1,p)}
     \overline{A}^{i+k}_{l-1,p}\sqrt{l} \right)\nonumber \\
 &&\mbox{Tr}_{\cal H}|_{\rm domain}
  [D_{\bar{2}},O]=
  \makebox{}\hspace{-10mm}
  \sum_{n,m,n',m',i,j,i',j'}^{\infty} \hspace{-8mm}
  \overline{A}^{i'}_{n',m'}S_{i'j'}S^{\dagger}_{ij}
  A^j_{n,m}O_{n,m,n',m'} \label{g-24} \\
 && \times \left( \sum_{l,p} A^{j'+k}_{l,p+1}
     \frac{\Lambda^{\frac{1}{2}}(l,p+1)}{ \Lambda^{\frac{1}{2}}(l,p)}
     \overline{A}^{i+k}_{l,p}\sqrt{p+1}
     -A^{j'+k}_{l,p}
     \frac{\Lambda^{\frac{1}{2}}(l,p)}{ \Lambda^{\frac{1}{2}}(l,p-1)}
     \overline{A}^{i+k}_{l,p-1}\sqrt{p} \right)\nonumber
\end{eqnarray}
Here we use Eq.(\ref{g-8}).
When we calculate the instanton charge,
these formulas play essential roles similar
to the Stokes' theorem in the commutative case.

\subsection{Converge condition}

In this subsection, we discuss converge conditions
for the instanton solution and study the behavior
of the solution in the large $\bar{z}z$ limit.

When the solution is given by (\ref{EQ:projection}) and the
absolute value of the eigenvalue of the
$\beta_{\alpha}$ is larger than cut-off $N$, 
our calculation is not valid.
So we do not consider the case that the locations of the instantons 
are on the point at infinity.
Then we can always choose the cut-off $N$ to be large enough as, 
\begin{eqnarray}
 0\le|\beta_{\alpha}^e|<N,\hspace{5mm}
  0\le|B_{\alpha}^e|<\sqrt{\zeta}N,\hspace{5mm}
  {\rm for \ any \ eigenvalue \ of \ \beta_{\alpha} (B_\alpha)},
  \label{c-1}
\end{eqnarray}
where $\beta_{\alpha}^e$ and $B_{\alpha}^e $
 represent eigenvalues of $\beta_{\alpha}$ and $B_\alpha$.\\
Let us consider the behavior of the gauge connection
with the condition (\ref{c-1}).
Since the projector $P$ is given by Eq.(\ref{EQ:projection}),
$| P \cket{n}{m}|$ is exponential
dumped as $n$ or $m$ increases.
On the other hand,
$ S^{\dagger} S \cket{n}{m}=(1-P)\cket{n}{m}$
approaches $\cket{n}{m}$ and this implies
\begin{eqnarray}
\lim_{n \rightarrow \infty} K_{n,m,n',m'}=
 \lim_{m \rightarrow \infty} K_{n,m,n',m'}=\delta_{nn'} \delta_{mm'}
\label{c-2}
\end{eqnarray}
Eq.(\ref{c-2}) is important to the obtain instanton charge. If we claim
condition (\ref{c-1}), the difference between $K_{nmn'm}$ and $\delta
_{nn'}\delta _{mm'}$ approach 0 as exponential function. But this is
too strong condition in practice. To estimate the instanton charge in
the next subsection , $K_{nmn'm}-\delta_{nn'}\delta
_{mm'}=O(1/(n^2+m^2)^2)$ is a necessary condition. 
This means that we can define the instanton charge as a converge series
for even in the case contains instantons on the point at infinity.
Because,  if we choose the location of the instantons with the cut-off $N$
whose radius from the origin is divergent in the large $N$ limit but
is enough slower divergent than the cut-off $N$,
then the instanton charge is well defined for the instanton that is
on the point of infinity in the large $N$ limit.\\

ASD Eqs.(\ref{g-11})
and (\ref{g-13}) are consistent with these conditions. The
$\Lambda({n_1,n_2})$
behavior is given by Eq.(\ref{EQ:Lambda-behavior}) in section
\ref{SEC:elongated}. Therefore, in the limit that one of $n,m,n'$ and
$m'$ approaches $\infty$, the ASD Eqs.(\ref{g-11}) and (\ref{g-13})
change to
\begin{eqnarray}
 &&K_{n'+1,m',n,m-1}\sqrt{(n'+1)m}=K_{n',m'+1,n-1,m}\sqrt{(m'+1)n},\\
 &&2K_{l',p',l,p}=K_{l',p',n',m'}K_{n,m,l,p}\nonumber\\
 &&\hspace{20mm}\times\left\{K_{n'+1,m',n+1,m}\sqrt{(n'+1)(n+1)}
         -K_{n'-1,m',n-1,m}\sqrt{n'n}\right.\nonumber\\
 &&\hspace{25mm}\left.+K_{n',m'+1,n,m+1}\sqrt{(m'+1)(m+1)}
   -K_{n',m'-1,n,m-1}\sqrt{m'm}\right\},
\end{eqnarray}
and Eq.(\ref{c-2}) obeys them.
These conditions are natural because if number is large enough
then the effect from instantons that locate in the are of
radius $\sqrt{\zeta}$ do not reach and the gauge field approaches
to the trivial connection.\\
As a result of Eq.(\ref{c-2}), we will see soon that
 calculation of instanton charge
become very simple with the Stokes' theorem
of the $D_{\bar{\alpha}}$.

The preparations to calculate the instanton charge
for general cases are complete.
In the next subsection, we start the calculation.

\subsection{Instanton number : general case}
Let us carry out the calculation of the instanton charge
with only primitive methods.
The instanton number is written as
\begin{eqnarray}
 Q&=&-\mbox{Tr}_{\cal H}1+\frac{\zeta^2}{2}\mbox{Tr}_{\cal H}
  \left\{
   [D_1,D_{\bar{1}}][D_2,D_{\bar{2}}] +[D_2,D_{\bar{2}}][D_1,D_{\bar{1}}]
        \right. \label{q-1}\\
 &&\left.\hspace{2.5cm}-[D_1,D_{\bar{2}}][D_2,D_{\bar{1}}]
    -[D_2,D_{\bar{1}}][D_1,D_{\bar{2}}] \right\}\nonumber\\
 &=&-\mbox{Tr}_{\cal H}1+\frac{\zeta^2}{2}\mbox{Tr}_{\cal H}
  \left\{ [D_{\bar{2}},\ D_2 D_{\bar{1}} D_1-D_1 D_{\bar{1}}D_2]
   +[D_{\bar{1}},\ D_1 D_{\bar{2}} D_2-D_2 D_{\bar{2}}D_1] \right\},
   \label{q-1a}\nonumber
\end{eqnarray}
where we use the ASD Eqs.(\ref{g-10}) and (\ref{g-12}).
In the following, we introduce a cut off $N$ like the previous
section and
the trace is defined with the cut off $N$.

\begin{figure}[t]
  \begin{center}
   \psfragscanon
   \psfrag{n1}[][][1.5]{$n_1$}
   \psfrag{n2}[][][1.5]{$n_2$}
   \psfrag{c1}[][][1.5]{$N$}
   \psfrag{c2}[][][1.5]{$N$}
   \psfrag{0} [][][1.5]{$0$}
   \psfrag{m} [][][1.5]{$m$}
   \psfrag{l} [][][1.5]{$l$}
   \psfrag{nm}[][][1.5]{$N_1(m)$}
   \psfrag{nl}[][][1.5]{$N_2(l)$}
   \scalebox{0.5}{\includegraphics{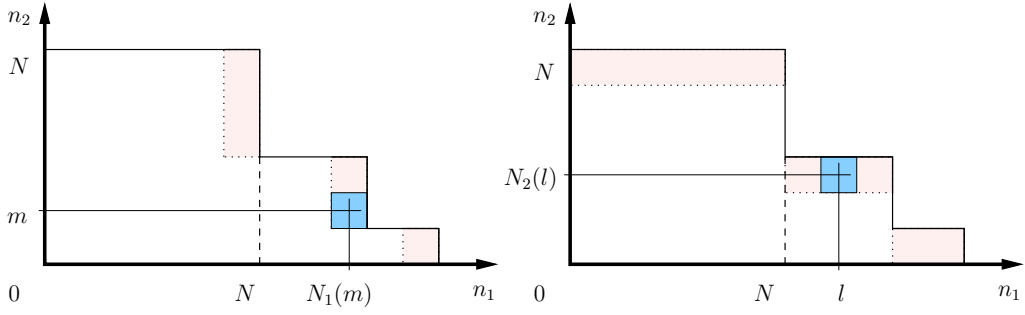}}
  \end{center}
 \caption{$N_1(m)$, $N_2(l)$; The set of the Fock states is composed of
 $(N+1)\times (N+1)$ square and extra $k$ cells.
 The extra $k$ cells a kind of Young-diagram and they are attached
 to $n_1 (n_2)$ axis and the square.
 $N_1(N_2)$ depend on $n_1(n_2)$.}
 \label{FIG:N1m-N2l}
\end{figure}

There are some points that we must make clarify.
If we defined the trace of $1$
by the $\sum_{n,m=0}^N 1$, how can we take the sum of
intermediate process?
As an example, let us consider the trace of $S^{\dagger}S$.
Remind that we choose the orthogonal basis $A^i$ as Eq.(\ref{g-7}).
When the cut off is given by $N$ we have to
prepare $(N+1)^2 +k$ Fock space basis, because the
$S$ and $S^{\dagger}$ are defined by $A^i$ and $A^{i+k}$
for $i=1,\cdots ,(N+1)^2$.
Using the convergent condition in the previous
subsection, orthogonal condition (\ref{c-2}) is realized for sufficient
large $N$,
then $(N+1)^2 +k$ Fock states is possible to
 expand the $(N+1)^2 +k$ orthonormal
basis $A^i$. In this situation, the trace of  $S^{\dagger}S$ is written
explicitly as
\begin{eqnarray}
    \mbox{Tr}_{\cal H}
  |_{\scsc {\rm cut\ off}}
  S^{\dagger}S
      &=&
        \sum_{n,m=0}^{{\rm cut~off}\ N}
         \sum_{i,j,p}
          \overline{A}^{i+k}_{nm}
        S^{\dagger}_{ij} S_{jp} A^{p+k}_{nm}
            \nonumber \\
 &=&
              \sum_{i,j}^{N} S^{\dagger}_{ij} S_{ji}.
               \label{q-1.1}
\end{eqnarray}
It is possible that we take $S_{ij}$ as a unitary matrices.
(As we saw at the beginning of this section, though $S_{ij}$ with the cut
off
$N$ is a unitary matrix $U(N+1)$, $S$ is still a partial
isometry operator.) Consequently, Eq.(\ref{q-1a}) become as follows:
\begin{eqnarray}
        \mbox{Tr}_{\cal H}
     |_{{\rm cut\ off}}
      S^{\dagger}S
    =
            \mbox{Tr}_{\cal H}
         |_{{\rm cut\ off}}{\bf 1}
\end{eqnarray}
Someone might think this equation is contradict to
$ S^{\dagger}S = 1-P$ .
However, infinite series like our instanton charge is sometime
conditionally convergent and
in such case, we have to introduce a cut off
and calculation manner.
If we do not provide $(N+1)^2+k$ Fock states but $(N+1)^2$,
then $\sum_{n,m}^N \bra{n}{m} S^{\dagger}S \cket{n}{m}$ is equal to
$(N+1)^2 -k$.
We can perform the calculation in the both ways,
but the later is little complex because we have to
estimate the lack of Fock space basis.
Therefore, in this study we choose the way
to provide  $(N+1)^2+k$ basis. For simplicity
we take the domain for the sum as (Fig.\ref{FIG:N1m-N2l}).
It is composite of a $(N+1) \times (N+1)$ square and  $k$ cells outside
of the square. We can put extra $k$ cells everywhere without inside of
the square. However, when the Stokes' theorem is used, it is convenience  to
take $k$ cells boundary like (Fig.\ref{FIG:N1m-N2l}).
As we saw in the previous subsection, Stokes' theorem change the
summation over domain to its on boundary.
If we set $k$ cells configuration arbitrary
, we have to take account each cells of extra $k$ cells
separately into the following calculation. In the following instanton
charge calculation, the difference arisen from the configuration of $k$
cells
is easily estimated as $O(k/N^2)$. ($O(k/N^2)$ is obtained from
$h_{\alpha}$ linear sum over $k$ points and leading term is not depend
on the $k$ cells locations.) We can check this by to perform
the following calculation with
arbitrary configuration of $k$ cells.
Therefore the instanton number do not depend on the locations of these
cells in the large $N$ limit and we perform the calculation with only
the domain of (Fig.\ref{FIG:N1m-N2l}) for simplicity.\\

The calculation of instanton number is primitive but
sometime complex, so we show the calculation explicitly.
For convenience, we introduce $C^{ \alpha\bar{\beta}\beta}$ as
\begin{equation}
 D_{\alpha}D_{\bar{\beta}}D_{\beta}
  =-\sqrt{\frac{1}{\zeta^3}}SC^{ \alpha\bar{\beta}\beta}S^{\dagger},
\end{equation}
where
\begin{eqnarray}
C^{ \alpha\bar{\beta}\beta}_{n,m,n',m'} \cket{n}{m}\bra{n'}{m'}
\!\!\!\!&=& \!\!\!\!
\Lambda^{-\frac{1}{2}}(\htn_1,\htn_2) c_{\alpha}
\Lambda^{\frac{1}{2}}(\htn_1,\htn_2) (1-P)
\Lambda^{\frac{1}{2}}(\htn_1,\htn_2) \\
&&\!\!\!\! \times \ c^{\dagger}_{\beta}
\Lambda^{-\frac{1}{2}}(\htn_1,\htn_2) (1-P)
\Lambda^{-\frac{1}{2}}(\htn_1,\htn_2) c_{\beta}
\Lambda^{\frac{1}{2}}(\htn_1,\htn_2) . \nonumber
\end{eqnarray}
The formula (\ref{g-24}) tells us that the instanton number
is given by a sum of progressions of difference and it become the
boundary integral. Indeed, explicit expression of
$\mbox{Tr}_{\cal H}|_{{\rm cut\ off}}[D_{\bar{2}},D_2D_{\bar{1}}D_1]$ is
\begin{eqnarray}
 &&\mbox{Tr}_{\cal H}|_{{\rm cut\ off}}[D_{\bar{2}},D_2D_{\bar{1}}D_1]
  \nonumber\\
 &&=-\frac{1}{\zeta^2}\sum_{{\rm all \ indices}}^{{\rm cut\ off}\ N}
  \left\{
   K_{l,p,n,m}K_{n',m',l,p+1}C^{1\bar{2}2}_{n,m,n',m'} \sqrt{p+1}
   \frac{\Lambda^{\frac{1}{2}}(l,p+1)}{\Lambda^{\frac{1}{2}}(l,p)}
  \right.\nonumber\\
 &&\hspace{30mm}
  \left.-K_{l,p-1,n,m}K_{n',m',l,p}C^{1\bar{2}2}_{n,m,n',m'}\sqrt{p}
   \frac{\Lambda^{\frac{1}{2}}(l,p)}{\Lambda^{\frac{1}{2}}(l,p-1)}
  \right\} \label{EQ:96}\\
 &&=-\frac{1}{\zeta^2}\sum_{{\rm all \ indices}}^{{\rm cut\ off}\ N}
  \left.\left\{
   K_{l,N,n,m}K_{n',m',l,N+1}C^{1\bar{2}2}_{n,m,n',m'} \sqrt{N+1}
   \frac{\Lambda^{\frac{1}{2}}(l,N+1)}{\Lambda^{\frac{1}{2}}(l,N)}\right\}
   \right|_{N=N_2(l)},\nonumber
\end{eqnarray}
where the cut off $N_2$ depend on $l$ as (Fig.\ref{FIG:N1m-N2l}).

When we claim the converge condition in previous subsection and we take
$N$ large enough, $K_{n,m,n',m'}$ is possible to be substituted by
$\delta_{nm}
\delta_{n'm'}$ in Eq.(\ref{EQ:96}).
Then the instanton charge is given by the following simple form:
\begin{eqnarray}
 &&
    Q=-\sum_{\rm square}1 \\
 &&
        -\sum_{n}\sqrt{N+1}
         \left. \left\{
       C^{2\bar{1}1}_{n,N,n,N+1}
            \frac{\Lambda^{\frac{1}{2}}(n,N+1)}
                  {\Lambda^{\frac{1}{2}}(n,N)}
         -C^{1\bar{1}2}_{n,N,n,N+1}
               \frac{\Lambda^{\frac{1}{2}}(n,N+1)}
               {\Lambda^{\frac{1}{2}}(n,N)})
                 \right\}\right|_{N=N_2(n)}\nonumber\\
 &&
        -\sum_{m}\sqrt{N+1}
          \left.\left\{
       C^{1\bar{2}2}_{N,m,N+1,m}
            \frac{\Lambda^{\frac{1}{2}}(N+1,m)}
                  {\Lambda^{\frac{1}{2}}(N,m)}
         -C^{2\bar{2}1}_{N,m,N+1,m}
               \frac{\Lambda^{\frac{1}{2}}(N+1,m)}
               {\Lambda^{\frac{1}{2}}(N,m)}
                 \right\}\right|_{N=N_1(m)},\nonumber
\end{eqnarray}
where we denote $N=N_{\alpha}(n)$ as the boundary of the domain
(Fig.\ref{FIG:N1m-N2l}).
Note that the sum of the first term is defined on the only square
without extra $k$ cells  i.e. it is
equal to  $(N+1)^2$. But the other sum is defined on the square and
the extra $k$ cells because the $S$ and $S^{\dagger}$ project out
$k$ dimensional subspace from the square.
Concrete expression of $C$ for large $N$ is, for example,
\begin{eqnarray}
 C^{1\bar{2}2}_{N,m,N+1,m}= \sqrt{(N+1)m^2}
  \frac{\Lambda^{\frac{1}{2}}(N+1,m)}{\Lambda^{\frac{1}{2}}(N,m)}
  \frac{\Lambda(N+1,m)}{\Lambda(N+1,m-1)}. \label{q-5}
\end{eqnarray}
Using Eq.(\ref{q-5}) and so on, $Q$ is
\begin{eqnarray}
 &&
      Q=-\sum_{\rm square}1
          \\
 &&
       \hspace{10mm}
       -\frac{1}{2}\sum_n
         \left\{
    (N+1)n\left(1-h_1(n-1,N+1)\right)
           \left(1-h_2(n,N)\right)
            \right.
             \nonumber\\
 &&
            \hspace{20mm}
             \left.
   -(N+1)(n+1)\left(1-h_1(n,N)\right)\left(1-h_2(n,N)\right)
               \right\}|_{N=N_2(n)}
                \nonumber\\
 &&
         \hspace{10mm}
         -\frac{1}{2}\sum_m
           \left\{
      (N+1)m\left(1-h_2(N+1,m-1)\right)
             \left(1-h_1(N,m)\right)
              \right.\nonumber\\
 &&
             \hspace{20mm}
              \left.
    -(N+1)(m+1)\left(1-h_2(N,m)\right)
                \left(1-h_1(N,m)\right)
                 \right\}|_{N=N_1(n)},
                  \nonumber
\end{eqnarray}
where we use $h_{\alpha}(n,m)$ defined in Eq.(\ref{EQ:h-definition}).
Since $\Lambda(n+1,m+1)/\Lambda(n,m)$
is written as
$(\Lambda(n+1,m)/\Lambda(n,m))(\Lambda(n+1,m+1)/\Lambda(n+1,m))$
or $(\Lambda(n,m+1)/\Lambda(n,m))(\Lambda(n+1,m+1)/\Lambda(n,m+1))$,
$h_\alpha$ obeys following consistency condition,
\begin{eqnarray}
h_1(n,m)+h_2(n+1,m)=h_2(n,m)+h_1(n,m+1)+O(\tfrac{1}{(n^2+m^2)^2}),
\label{q-7}
\end{eqnarray}
where $O(\tfrac{1}{(n^2+m^2)^2})$ is the second power of $h_{\alpha}$.
Using this Eq.(\ref{q-7}), up to $O(\tfrac{1}{N})$ the instanton charge
$Q$ is written as follows:
\begin{eqnarray}
      Q
 &=&
     -\sum_{\rm square}1
      +\frac{1}{2}\sum_n(N+1)|_{N=N_2(n)}
       +\frac{1}{2}\sum_m(N+1)|_{N=N_1(m)}
        \label{q-8} \\
 &&
        -\sum_n (N+1)h_2(n,N)|_{N=N_2(n)}
         -\sum_m (N+1)h_2(N,m)|_{N=N_1(m)}
           \nonumber \\
 &&
        -\frac{1}{2}\sum_n
          \left\{ (N+1)\left( (n+1)h_1(n,N)-n\;h_1(n-1,N)\right)
           \right.
            \nonumber \\
 &&
       \left.
  -(N+1)\left( (n+1)h_2(n,N)-n\;h_2(n-1,N)
         \right) \right\}|_{N=N_2(n)}
          \nonumber \\
 &&
     -\frac{1}{2}\sum_m
       \left\{ (N+1)\left( (m+1)h_2(N,m)-m\;h_2(N,m-1)\right)
        \right.
         \nonumber \\
 &&
           \left.
      -(N+1)\left( (m+1)h_1(N,m)-m\;h_1(N,m-1)
             \right) \right\}
            |_{N=N_1(m)}.
               \nonumber
\end{eqnarray}
The last four lines of (\ref{q-8})
 are 0 up to $O(k(N+1)h_{\alpha})\sim O(\frac{k}{N})$.

Taking a trace ${\rm Tr}_{\scsc [0,~N]}$ of Eq.(\ref{g-12}), following
equation is obtained:
\begin{eqnarray}
 2\sum_{\rm square}1&=&\sum_m(N+1)(1-h_1(N,m))|_{N=N_1(m)}\nonumber\\
 &&+\sum_n(N+1)(1-h_2(n,N))|_{N=N_2(n)}+O(\tfrac{k}{N}).
\end{eqnarray}
$\sum (N+1)|_{N=N_\alpha}$ is equal to
 $(N+1)^2 +k$ for both ${\alpha=1}$ and 2.
Hence, we finally get the instanton charge as
\begin{eqnarray}
 Q_{\scsc {\rm cut \hspace{1mm} off}}=-k+O(\tfrac{k}{N}),
\end{eqnarray}
and it is equivalent to the instanton number in the large $N$ limit.

\section{Conclusion and Summary}
We have studied the analytical derivation of the instanton number as
the integral of
the Pontrjagin class (instanton charge) in the Fock space representation.
Our approach was for the general noncommutative $U(1)$ instanton solution,
and was based on
the stability condition and the anti-self dual equation itself. At first
the instanton charge of an elongated type was calculated.(In the
paper~\cite{Ishikawa1},
we constructed the noncommutative elongated instanton solution,
and performed the instanton charge calculus by the numerical calculation.)
After that, general instanton calculation was done by similar manner.
To avoid ambiguity, we introduced the cut-off instanton charge
and the boundary.
All the calculations were done in finite and all the
surface terms were estimated.
Additionally, all the calculations were performed with algebraic methods.
These results shows that the direct relation between the differential
geometry and recurrence relations and its series.

Rough sketch of our calculation is following. We made the shift operator
with the orthonormal basis. This shift operator played essential role.
We gave an orthonormalization which decomposes into the $k$ eigenvectors
of projector $P$
 and its orthogonal complement. The projector was given by  general form,
including, e.g., coherent states. We gave the Stokes' theorem for the
number operator representation. The Stokes' theorem on the
noncommutative space shows that a trace over arbitrary regions of a
commutator with coordinate operators is translated into some boundary
sum. The commutator with coordinate operators is a derivative, and is
a finite-difference form. As same as the commutator with the gauge
connection behaves a finite-difference form when we take its trace.
For this feature, the trace operation in the
instanton charge calculation became the sum over the boundary.
Therefore the instanton charge calculus almost came back to
elongated type calculation. From this realization we provided the
instanton number up to $O(k/N)$, and in large $N$ limit this difference
vanish.
Our calculation clarified that
the origin of the instanton charge is the dimension of the projection.
In other words, the projector in ADHM construction directly
reflects the projection in noncommutative field theory.

When we survey our calculation, someone might think the instanton charge
is conditionally converge. However it is hasty, because we can
define the instanton charge as the sum of each density of each Fock state
including constant curvature. 
If we do not introduce the cut-off then the instanton charge is
conditionally convergence in general and its value is meaningless.
But we define the instanton charge as $\lim_{N \rightarrow \infty} Q_N$
with cut-off $N$.
This $Q_N$ is a converge series since the difference between instanton
charge and instanton number is $O(k/N)$ and it is monotone decreasing
along increasing $N$.
On the other hand, the cut-off brings about a new problem that we have to
check the boundary(cut-off) dependence of the instanton charge.
We are able to verify following facts.
The instanton charge is not changed when we choose the cut-off
not as $N \times N$ square but
as $N_1 \times N_2$ rectangle or triangle that is defined by
$n_1 + n_2 = N$.
In the rectangle boundary case, the  difference between instanton charge
and instanton number is $O(k/N_1 )+O(k/N_2)$ and it is monotone decreasing
along both $n_1$ and $n_2$ directions.
The triangle boundary case is similar too.
Further the instanton charge is invariant under
the finite deformation of the boundary.
Therefore, it is natural to understand that the instanton charge is defined
by a converge series whose accumulation value do not depend on the detail of
the boundary.\\
We expect that the extension of this work to
a $U(N)$ case is not difficult to do.
Furthermore the other type of
characteristic class might be calculated as the same manner.
These subjects are left for future works.


\section*{Acknowledgements}

We would like to thank Mikio Furuta, Yoshiaki Maeda, and Hitoshi
Moriyoshi for useful discussion. A.S is supported by JSPS Research
Fellowships for Young Scientists. Discussions during the YITP workshop
YITP-W-01-04 ``Quantum Field Theory 2001'' were helpful to complete this
work.

\newpage

\appendix

\section{Calculation of instanton charge : elongated $U(1)$ instantons}
\label{AP:charge-elongated}

In this appendix we show the detail of the calculation of the instanton
charge $Q$ in section \ref{SEC:elongated-number}.\\

$Q$ is possible to be written with progression of differences:
\begin{eqnarray}
 &&Q_{\rm cutoff}\nonumber\\
 &&=\zeta^2\mbox{Tr}_{\cal H}\big|_{\scsc [0,~N]}
  \left\{F_{1\bar{1}}F_{2\bar{2}}
   -\frac{1}{2}(F_{1\bar{2}}F_{2\bar{1}}+F_{2\bar{1}}F_{1\bar{2}})
  \right\}\nonumber\\
 &&=\mbox{Tr}_{\cal H}\big|_{\scsc [0,~N]}\biggl\{-1
  +\zeta^2[D_1,D_{\bar{1}}][D_2,D_{\bar 2}]
  -\frac{\zeta^2}{2}\bigl([D_1,D_{\bar{2}}][D_2,D_{\bar{1}}]
  +[D_2,D_{\bar{1}}][D_1,D_{\bar{2}}]\bigr)\biggr\}\nonumber\\
 &&=\sum_{(n_1,n_2)}(-1)\times\theta\bigl((n_1,n_2)\in R_0\bigr)
  +\sum_{(n_1,n_2)}1\times\theta\bigl((n_1,n_2)\in R_S\bigr)\nonumber\\
 &&-\frac{1}{2}\sum_{(n_1,n_2)}\biggl[
  (n_1+1)\Bigl\{
  (n_2+1)\bigl(h_1(n_1,n_2)+h_2(n_1,n_2)-h_1(n_1,n_2)h_2(n_1,n_2)\bigr)
  \nonumber\\
 &&\hspace{20mm}
  -n_2\bigl(h_1(n_1,n_2-1)+h_2(n_1,n_2-1)-h_1(n_1,n_2-1)h_2(n_1,n_2-1)\bigr)
  \Bigr\}\nonumber\\
 &&-n_1\Bigl\{
  (n_2+1)\bigl(h_1(n_1-1,n_2+1)+h_2(n_1,n_2)-h_1(n_1-1,n_2+1)h_2(n_1,n_2)
  \bigr)\nonumber\\
 &&\hspace{20mm}
  -n_2\bigl(h_1(n_1-1,n_2)+h_2(n_1,n_2-1)-h_1(n_1-1,n_2)h_2(n_1,n_2-1)\bigr)
  \Bigr\}\nonumber\\
 &&+(n_2+1)\Bigl\{
  (n_1+1)\bigl(h_1(n_1,n_2)+h_2(n_1,n_2)-h_1(n_1,n_2)h_2(n_1,n_2)\bigr)
  \nonumber\\
 &&\hspace{20mm}
  -n_1\bigl(h_1(n_1-1,n_2)+h_2(n_1-1,n_2)-h_1(n_1-1,n_2)h_2(n_1-1,n_2)\bigr)
  \Bigr\}\nonumber\\
 &&-n_2\Bigl\{
  (n_1+1)\bigl(h_1(n_1,n_2)+h_2(n_1+1,n_2-1)-h_1(n_1,n_2)h_2(n_1+1,n_2-1)
  \bigr)\nonumber\\
 &&\hspace{20mm}
  -n_1\bigl(h_1(n_1-1,n_2)+h_2(n_1,n_2-1)-h_1(n_1-1,n_2)h_2(n_1,n_2-1)\bigr)
  \Bigr\}\biggr]\nonumber\\
 &&\hspace{100mm}\times\theta\bigl((n_1,n_2)\in R_S\bigr)\;.\nonumber\\
\end{eqnarray}
The first term and the second term are canceled out.
The remaining terms are written by the form of difference, so we obtain
the summation only on the cut-off and the result is
\begin{eqnarray}
 &&Q_{\rm cutoff}=k\nonumber\\
 &&-\frac{1}{2}\biggl[
 (N+1)\sum_{n_1=0}^N\Bigl\{
 (n_1+1)\bigl(h_1(n_1,N)+h_2(n_1,N)-h_1(n_1,N)h_2(n_1,N)\bigr)\nonumber\\
 &&\hspace{25mm}
  -n_1\bigl(h_1(n_1-1,N+1)+h_2(n_1,N)-h_1(n_1-1,N+1)h_2(n_1,N)\bigr)\Bigr\}
  \nonumber\\
 &&\hspace{5mm}
  +\sum_{n_1=N+1}^{N+k}\Bigl\{
  (n_1+1)\bigl(h_1(n_1,0)+h_2(n_1,0)-h_1(n_1,0)h_2(n_1,0)\bigr)\nonumber\\
 &&\hspace{25mm}
  -n_1\bigl(h_1(n_1-1,0)+h_2(n_1,0)-h_1(n_1-1,0)h_2(n_1,0)\bigr)\Bigr\}
  \nonumber\\
 &&\hspace{5mm}
  +(N+k+1)\bigl(h_1(N+k,0)+h_2(N+k,0)-h_1(N+k,0)h_2(N+k,0)\bigr)\nonumber\\
 &&\hspace{5mm}
  +\sum_{n_2=1}^N(N+1)\Bigl\{
  (n_2+1)\bigl(h_1(N,n_2)+h_2(N,n_2)-h_1(N,n_2)h_2(N,n_2)\bigr)\nonumber\\
 &&\hspace{25mm}
  -n_2\bigl(h_1(N,n_2)+h_2(N+1,n_2-1)-h_1(N,n_2)h_2(N+1,n_2-1)\bigr)\Bigr\}
  \biggr].\nonumber\\
\end{eqnarray}
Using the relation (\ref{EQ:h-relation1}), we obtain
the form of difference again:
\begin{eqnarray}
 &&Q_{\rm cutoff}=k\nonumber\\
 &&-\frac{1}{2}\biggl[(N+1)\sum_{n_1=0}^N
  \Bigl\{(n_1+1)\bigl(h_1(n_1,N)-h_2(n_1,N)\bigr)\nonumber\\
 &&\hspace{40mm}
  -n_1\bigl(h_1(n_1-1,N)-h_2(n_1-1,N)\bigr)+2h_2(n_1,N)\Bigr\}\nonumber\\
 &&\hspace{10mm}
  +\sum_{n_1=N+1}^{N+k}\Bigl\{
  (n_1+1)\bigl(h_1(n_1,0)-h_2(n_1,0)\bigr)\nonumber\\
 &&\hspace{40mm}
  -n_1\bigl(h_1(n_1-1,0)-h_2(n_1-1,0)\bigr)+2h_2(n_1,0)\Bigr\}\nonumber\\
 &&\hspace{10mm}
  +(N+k+1)\bigl(h_1(N+k,0)+h_2(N+k,9)\bigr)\nonumber\\
 &&\hspace{10mm}
  -(N+1)\sum_{n_2=1}^N\Bigl\{
  (n_2+1)\bigl(h_1(N,n_2)-h_2(N,n_2)\bigr)\nonumber\\
 &&\hspace{40mm}
  -n_2\bigl(h_1(N,n_2-1)+h_2(N,n_2-1)\bigr)-2h_1(N,n_2)\Bigr\}
  \nonumber
\end{eqnarray}
\begin{eqnarray}
 &&+\frac{1}{2}\biggl[(N+1)\sum_{n_1=0}^{N}\Bigl\{\bigl(
  (n_1+1)h_1(n_1,N)-n_1h_2(n_1-1,N)\bigr)h_2(n_1,N)\nonumber\\
 &&\hspace{40mm}
  +n_1h_1(n_1-1,N+1)h_2(n_1-1,N)\Bigr\}\biggr]\nonumber\\
 &&\hspace{10mm}
  +(N+1)\sum_{n_2=1}^{N}\Bigl\{
  h_1(N,n_2)\bigl((n_2+1)h_2(N,n_2)-n_2h_2(N,n_2-1)\bigr)\nonumber\\
 &&\hspace{40mm}
  +n_2h_1(N,n_2-1)h_2(N+1,n_2-1)\Bigr\}\nonumber\\
 &&\hspace{10mm}
  +\sum_{n_1=N+1}^{N+k}\Bigl\{\bigl(
  (n_1+1)h_1(n_1,0)-n_1h_1(n_1-1,0)\bigr)h_2(n_1,0)\nonumber\\
 &&\hspace{40mm}+n_1h_1(n_1-1,1)h_2(n_1-1,0)\Bigr\}\nonumber\\
 &&\hspace{10mm}+(N+k+1)h_1(N+k,0)h_2(N+k,0)\biggr]\;.
\end{eqnarray}
We perform the summation and get
\begin{eqnarray}
 Q_{\rm cutoff}
 &=&k-\Bigl\{(N+1)\sum_{n_1=0}^Nh_2(n_1,N)
  +(N+1)\sum_{n_2=0}^Nh_1(N,n_2)\nonumber\\
 &&+\sum_{n_1=N+1}^{N+k}h_2(n_1,0)+(N+k+1)h_1(N+k,0)\Bigr\}\nonumber\\
 &&+(h_{\alpha}^2 \;\mbox{terms}).
\end{eqnarray}
The relation (\ref{EQ:E-equation}) shows that
\begin{equation}
 Q_{\rm cutoff}=k-E+(h_{\alpha}^2 \;\mbox{terms})
  =-k+(h_{\alpha}^2 \;\mbox{terms}).
\end{equation}


\end{document}